\documentclass[12pt,a4paper]{article}
\usepackage[margin=1in]{geometry}
\usepackage{amsmath,amsthm,amssymb,mathabx}
\usepackage[utf8]{inputenc}
\usepackage{rotating}
\usepackage{tikz}
\usepackage{pgfplots}
\usepackage{booktabs}
\usepackage{multirow}
\usepackage{subfig}
\usepackage{setspace}
\usepackage{enumitem}   
\usepackage{natbib}

\bibpunct{(}{)}{;}{a}{,}{,} 
\makeatletter
\let\ref\@refstar
\makeatother
\usetikzlibrary{shapes}
\pgfplotsset{compat=1.18}

\newtheorem{theorem}{Theorem}
\newtheorem{proposition}{Proposition}
\newtheorem{lemma}{Lemma}

\newtheorem{remark}{Remark}

\theoremstyle{definition}

\newtheorem{example}{Example}

\newcommand{\flatbid}{constant}
\newcommand{\Flatbid}{Constant}

\newcommand{\pif}{p_1^f}
\newcommand{\pjf}{p_2^f}
\newcommand{\pifull}{\frac{U_1(\lambda_1)-U_1(1-\lambda_2)}{\lambda_1}}
\newcommand{\pjfull}{\frac{U_2(\lambda_2)-U_2(1-\lambda_1)}{\lambda_2}}

\newcommand{\li}{\lambda_1}
\newcommand{\lj}{\lambda_2}

\usepackage{hyperref}

\begin{document}

\title{The Combinatorial Multi-Round Ascending Auction\thanks{We are very grateful to Larry Ausubel, Ravi Jagadeesan, Maarten Janssen, Simon Jantschgi, Paul Klemperer, Christian Koboldt, Dan Maldoom, Daniel Marszalec, Heinrich Nax, Marek Pycia, Roger Salsas, and Kyle Woodward for their comments, to an anonymous referee during an earlier submission for suggesting that we consider asymmetric caps, and to Edwin Lock for outstanding research assistance. We also thank audiences at the 2024 Tinbergen Workshop on Auctions \& Market Design (Amsterdam), 2023 SITE Market Design conference (Stanford), the Virtual Market Design Seminar, the 2022 INFORMS Revenue Management and Pricing Section Conference, and the 12th Conference on Economic Design (Padova) as well as in Berlin, Cologne, Düsseldorf, Manchester, Oxford, and Zurich for valuable feedback and comments. While it could be argued that the auction's name ought to be abbreviated to ``CMRAA,'' we follow the designers' original acronym.
}}
    
\author{Bernhard Kasberger\thanks{Department of Economics, Johannes Kepler University Linz; Austrian Institute of Economic Research, Email: \texttt{bernhard.kasberger@jku.at}} \and Alexander Teytelboym\thanks{Department of Economics and St. Catherine's College, University of Oxford. Email: \texttt{alexander.teytelboym@economics.ox.ac.uk}. This work was supported by the Economic and Social Research Council grant number ES/R007470/1 and has received funding from the European Research Council (ERC) under the European Union’s Horizon 2020 research and innovation programme (grant agreement No.~949699).}}

\maketitle
\begin{abstract}
The Combinatorial Multi-Round Ascending Auction (CMRA) is a new auction format used in recent European spectrum auctions. We show that an auction-specific version of truthful bidding leads to an efficient allocation. We then characterize different ex-post equilibria that feature truthful bidding, demand expansion, and demand reduction. The truthtelling equilibrium is fragile to small asymmetries in the bidders' caps. Moreover, if bidders are sufficiently symmetric, the CMRA is vulnerable to risk-free collusion. We propose an alternative activity rule that prevents such collusive strategies while keeping other equilibria intact. We discuss outcomes of several Danish CMRAs in light of our equilibrium predictions.
\end{abstract}

\bigskip


\bigskip


\onehalfspacing
\section{Introduction}

Auction sales of radio spectrum have served as a powerful engine for auction format innovation. These high-stake sales often involve heterogeneous spectrum bands and require flexible and robust auction formats. For example, the Simultaneous Multi-Round Auction (SMRA), the Combinatorial Clock Auction (CCA), and the Deferred Acceptance Auction, all trace their origins to spectrum auctions.\footnote{The United States pioneered the SMRA and the DA Auction in 1994 and 2016 respectively; the CCA was used for the time in 2005 for spectrum sales in Trinidad.} Many dynamic auctions work well---they find efficient outcomes, give bidders decent incentives to bid truthfully, and are fairly robust to collusion---when the goods for sale are substitutes \citep{kelso1982job,milgrom2000putting}. However, in the presence of complements---which is very common in spectrum allocation---auction design becomes substantially more difficult. To help bidders avoid ending up with undesirable combinations of goods, many \emph{combinatorial} auction formats, such as the CCA and the sealed-bid combinatorial auction, allow bidders to express preferences over packages. 
In its 2016 spectrum auction, the Danish Energy Agency (DEA) pioneered the \emph{Combinatorial Multi-Round Ascending Auction} (CMRA) format developed by the UK-based consultancy DotEcon Ltd. to sell several spectrum blocks in the 1800 MHz band \citep{dotecon2016proposal}. Since then, the CMRA has been used twice again by the DEA and once by the Norwegian regulator (Nkom).\footnote{Norway was also the first country in the world to use a (first-price) combinatorial auction to sell spectrum in 2001.} Such new combinatorial auction formats for spectrum are rare \citep{palacios2021combinatorial} and provide a unique opportunity to apply economic theory in a complex, high-stakes, real-world environment.

This is the first paper to provide a game-theoretic analysis of the CMRA. We isolate key strategic incentives which allow us to identify substantial shortcomings of the current design and suggest ways to improve it.
The CMRA works as follows. There is a price clock for each good. In each clock round, bidders can report a \emph{headline demand} for a single package which is priced linearly at clock prices. The novel feature of the CMRA is that in each clock round, bidders can also submit \emph{additional} (package) \emph{bids} at or below the clock prices (and subject to an activity rule). As a result, the CMRA does not work like a standard ascending auction because even at high clock prices bidders can reveal demand at low prices using the additional bids. The auction ends when there exists a revenue-maximizing allocation in which \emph{exactly one bid from every bidder} is accepted.

There are two main differences between the CMRA and the CCA, arguably the most frequently used combinatorial auction format in spectrum sales: the CMRA has no supplementary sealed-bid round (while the CCA does) and the CMRA allows for non-linearly priced additional bids in the clock stage (while the CCA does not). In the presence of complements, some form of non-linear pricing might be necessary to clear the market \citep{bikhchandani2002package}. To this end, the CCA typically uses Vickrey–Clarke–Groves (VCG)-like pricing in the supplementary round. By contrast, the CMRA allows bidders to submit non-linear (pay-as-bid-priced) additional bids in the clock phase. Hence, the market can clear in the clock phase, making the supplementary phase unnecessary.\footnote{The combinatorial auction due to \citet{porter2003combinatorial} also features no supplementary round.} Another difference is that the (clock phase of the) CCA ends when there is no excess demand for any good whereas the CMRA ends when the auctioneer can accept a bid from each bidder in a revenue-maximizing allocation.

There are several reasons why one might prefer to use an auction format with some of the features of the CMRA in practice. First, the CCA's supplementary round and pricing rule create uncertainty about what the bidders will win and what they will pay. Avoiding the supplementary round can therefore help bidders discover prices more gradually and deal more easily with their budget constraints \citep{janssen2017budget}. Second, submitting  a large number of mutually exclusive bids in the supplementary round (in order to achieve the most preferred auction outcome) can be taxing for bidders. Indeed, bidders might prefer the ability to revise the packages they bid on as clock prices change. Since the CMRA elicits demand information gradually, one might hope to overcome the ``missing bids'' problem of bidders failing to bid on all the packages they find valuable and that has been observed in CCAs \citep{bichler2013efficiency}. Third, the VCG-based pricing rule of the CCA (but not the pricing rule of the CMRA) can provide bidders the possibility of (spitefully) raising the other bidders' payments, which can be at the detriment of efficient allocations \citep{levin2016properties,janssen2019clock}. 

We begin by analyzing non-strategic bidding in a general model of the CMRA with arbitrary numbers of bidders and (divisible or indivisible) goods.
We focus on a version of ``truthful'' bidding  
which was also explicitly proposed by the auction designers \citep{dotecon2016proposal}.
Specifically, we consider the 
bidding strategy (which we call \emph{CMRA-truthful} bidding) in which each bidder's headline demand is a myopically optimal bundle at the current clock prices and each additional bid is set to make the bidder indifferent between winning the package with that bid and winning the headline demand at clock prices.

Our first main result (Theorem~\ref{prop:CMRA truthful}) shows that CMRA-truthful bidding leads to an efficient allocation irrespective of the bidders' preferences and even in settings where a competitive equilibrium is not guaranteed to exist (e.g., in the presence of non-convex preferences). 
Additional bids play a key role in ensuring the efficiency of CMRA-truthful bidding: using truthful headline demands without additional bids (what we call 
\emph{clock-truthful} bidding) can fail to guide the CMRA to an efficient outcome. However, the presence of truthful additional bids in CMRA-truthful bidding causes the CMRA to terminate before reaching (linear) market-clearing prices thereby yielding lower revenue than in the competitive equilibrium outcome (Proposition~\ref{prop:revenue-competitive-equ}).

Our remaining results concern strategic bidding in the CMRA. 
To analyze strategic bidding, we turn to a simpler model of two bidders and a single divisible good (cf. \citet{levin2016properties}) in order to cleanly isolate key incentives. Bidders' marginal utilities for the good can be decreasing (capturing substitutes) or non-decreasing (capturing independent goods or complements). We allow for caps, which are common in spectrum auctions. Thus, no single bidder is allowed to win the entire supply. Our focus is on ex-post equilibria in ``proxy'' strategies, i.e., bidders delegate bidding to a proxy agent that bids only according to prevailing prices without taking into account the actions of other bidders \citep{ausubel2002ascending}.

When marginal values are non-decreasing, CMRA-truthful bidding is an ex-post equilibrium when the bidders' caps are exactly symmetric. (This continues to be the case with any number of bidders.) With (arbitrarily small) asymmetries in the bidders' caps, however, CMRA-truthful bidding is no longer an equilibrium (Theorem~\ref{theorem:increasing-MV-CMRA-truthful-expost}). CMRA-truthful bidding is not an equilibrium with decreasing marginal values due to free-riding and a threshold problem (Proposition~\ref{theorem:noeqm}). Moreover, clock-truthful bidding is never an equilibrium: with decreasing marginal values, bidders have demand reduction incentives as in other multi-unit auctions \citep{ausubel2014demand} and with non-decreasing marginal values the weaker bidder has an incentive to close the auction early by bidding on the units ``left over'' by the stronger bidder. 

Next, we identify a simple strategy that forms an ex-post equilibrium both under decreasing and non-decreasing marginal values (Theorem~\ref{theorem:constant-ex-post}). With symmetric caps, the (what we call \emph{constant}) strategy prescribes a constant headline demand for the maximum (capped) quantity and a single additional bid of zero (i.e., the reserve price) for the residual supply at a relatively high clock price. The strategy thereby resembles clock-truthful bidding under non-decreasing marginal values except for the single additional bid. In this equilibrium and with decreasing marginal values, the bidders expand demand so that the stronger bidder wins a larger quantity than what she would have won in the efficient allocation. With non-decreasing marginal values, the equilibrium outcome is as under CMRA-truthful bidding and also as in a VCG auction under truthful bidding. 
With asymmetric caps, both bidders' headline demand is the second highest cap and the ex-post equilibrium exists as long as the utility for quantities between the two caps is not too different.

A final and crucial theoretical insight for practical auction design is that bidders can use additional bids to collude. Suppose each bidder plays the constant strategy, but also places an additional bid of half the initial clock price for half of the available quantity in the initial clock round. As both bidders winning half of the supply is (weakly) revenue-maximizing, the auction ends immediately by assigning half of the supply to each bidder at the initial clock price. 
We show that this strategy profile forms an equilibrium if bidders are sufficiently symmetric (Theorem~\ref{theorem:collusion}). Surprisingly, this collusive demand reduction strategy carries \emph{no risk}. If the other bidder does not work out the collusive bidding strategy and the auction does not end in the first round, the bidder is unconstrained in her bids in later rounds. In any case, the weaker bidder loses nothing by placing a zero additional bid on half of the supply.

We then argue that the driving force behind the collusion incentives is the activity rule used in the CMRA. Activity rules typically prevent bidders from underreporting their demand early in the auction \citep{ausubel2020revealed,milgrom2000putting}. However, the activity rule currently used in the CMRA only transforms headline demands into constraints on the bid function. 
We propose an alternative activity rule that makes the risk-free demand reduction strategy infeasible: Our alternative activity rule treats additional bids and headline demands identically by transforming both into constraints on bidding. The alternative activity rule keeps the other equilibria we discussed intact; however, it might force the bidders to revise the bids on more packages throughout the auction.

Finally, we connect our theoretical predictions to outcomes of CMRAs in practice. We look at the outcomes of three Danish spectrum auctions held between 2016 and 2021. As the DEA only publishes data on allocations and total bidder payments, we try to reverse-engineer whether headline demands or additional bids were winning in the auctions. We then use these insights to speculate whether the possible bidding dynamics were consistent with our theoretical predictions. It appears that in the 2016 auction only headline demands won, which suggests that bidders behaved as if they were in a clock auction. In the 2019 and 2021 auctions, some additional bids were winning, suggesting that bidders were using much richer strategies in later auctions. However, we do not find any evidence that bidders use the risk-free collusive strategy that we outline. We also explain what our results mean for patterns of bidding dynamics and how our theoretical predictions could be tested on actual CMRA bidding data.

The literature on combinatorial auctions is vast.
Combinatorial auctions were introduced in the context of selling airport take-off and landing slots by \citet{rassenti1982combinatorial}. 
\cite{Bernheim-Whinston-Menu-Auction} introduced a static combinatorial (``menu'') auction which shares the pay-as-bid pricing rule with the CMRA's additional bids. Under complete information with two bidders, the ``menu'' auction admits a Nash equilibrium in strategies that resemble the CMRA-truthful strategy and features VCG prices. However, CMRA-truthful bidding is not always an equilibrium in the CMRA auction and only features VCG prices in special (symmetric) cases (see Section~\ref{sec:strategic}). Following the success of the SMRA for early US spectrum auctions in 1990s \citep{milgrom2000putting}, license complementarities in subsequent sales were substantial enough to warrant new auction formats. As a result, a number of combinatorial auction designs were proposed \citep{parkes2002ascending,porter2003combinatorial}.\footnote{ \citet{palacios2021combinatorial} offer an excellent recent overview of the use of combinatorial auctions, including CMRAs, in practice.}

A major breakthrough for practical auction design was the development of the CCA \citep{ausubel2006clock,maldoom2007winner} which quickly became one of the dominant auction formats for spectrum sales and beyond. The CCA was theoretically analyzed by \cite{levin2016properties}, \cite{ janssen2016spiteful} and \citet{janssen2019clock}. We use many modelling features from the elegant analysis of \citet{levin2016properties} as well as their focus on proxy strategies introduced by \citet{ausubel2002ascending}. Unlike many other papers, we consider both increasing and decreasing marginal values.

On the theoretical side, many papers have proposed innovative dynamic auction formats for homogeneous goods \citep{ausubel2004efficient}, heterogeneous substitutes \citep{ausubel2006AER}, heterogeneous complements \citep{sun2014efficient} as well as general valuations \citep{mishra2007ascending} in which truthful bidding is an ex-post equilibrium and final payments coincide with VCG prices.\footnote{\citet{sun2009double} and \citet{teytelboym2014gross} proposed dynamic auctions for selling restricted classes of substitutes and complements without offering a strategic analysis.} The ``interval bidding auction procedure'' proposed by \citet{baranov2017efficient} is particularly related to our setting because their ``interval bids'' play a similar role to CMRA's additional bids by allowing bidders to express preferences over complements. However, most of these formats are not used in practice. What makes the CMRA interesting to study is that it (i) has been used in practice; (ii) has an unusual format due the additional bids; (iii) admits a variety of ex-post equilibria, but (iv) has a truthful equilibrium only in certain cases.

A large literature has studied (collusive) demand reduction in static and dynamic multi-unit auctions (e.g., \citet{ausubel-schwartz,GRIMM20031557,ausubel2014demand}). The main difference is that in our case the demand reduction strategy is risk-free while in other auction formats it is not. We also observe demand \emph{expansion} in equilibrium, which, to the best of our knowledge, has not been noted as an equilibrium phenomenon in auctions with pay-as-bid pricing.

We proceed as follows. The next section illustrates the dynamics of the clock auction and the CMRA under different bidding strategies. Section~\ref{sec:model} presents the formal model and the auction rules. In Section~\ref{sec:truthful}, we analyze non-strategic truthful bidding and then turn to strategic bidding in Section~\ref{sec:strategic}. Section~\ref{sec:risk free} describes the risk-free collusive strategy and proposes an alternative activity rule to deal with it. Section~\ref{sec:realworld} explains how to use bidding data to test which bidding strategies might have been played. We describe how the CMRA performed in the Danish spectrum auctions in Section~\ref{sec:Danish auctions}. Section~\ref{sec:conclusion} is a conclusion. All omitted proofs are in Appendix~\ref{app:proofs}. The extension to the case of any number of bidders is Appendix~\ref{app: more than two bidders}. Further details about the Danish auctions and illustrations of CMRA-truthful bidding are in the supplemental appendix.

\section{Illustrative Example: Clock Auction vs. CMRA}\label{sec:example}
To make the CMRA more accessible, we present a simple example that demonstrates its rules and workings. As the CMRA builds on the clock auction, we first contrast the CMRA with the simplest possible clock auction to shed some light on the role of the CMRA's additional bids. As we explore the example further, we illustrate some of our main findings. 

There are four identical lots for sale and two bidders. Bidders are symmetric: The valuation of each lot is \$30 and the valuations are additive for each bidder. There is a symmetric cap of three lots for either bidder. 

In a (standard) clock auction, the price starts at \$0 and increases as long as there is excess demand. Under truthful bidding, each bidder demands three lots as long as the clock price is below \$30. At \$30 the clock stops because both bidders drop their demands to 0. The bidders have expressed demand for three units which can be allocated to either bidder, therefore there is excess supply of one lot. Hence, the final allocation is inefficient. The auction revenue is \$90.\footnote{Depending on the design of the clock auction, it is possible that when bidders drop their demand to zero simultaneously, the auctioneer allocates nothing to the bidders. In this case, imagine that the valuation of one of the bidders is \$30$+\epsilon$. Note that there are also installations of the clock auction that would allow market clearing.}

Next, consider how the CMRA would run. Bidders can mimic their strategies in the clock auction by only submitting a headline demand (and no additional bids) at each clock price. If the headline demands are truthful---this is the \emph{clock-truthful} setting---and the price is below \$30, then the revenue-maximizing allocation involves only \emph{one} bidder; therefore, the price continues to increase. At \$30, bidders drop their headline demand to zero lots. Now the revenue-maximizing allocation can involve a bid from both bidders: we can allocate three units to one bidder and no units to the other bidder. Therefore, the outcome is the same as in the clock auction. It is easy to see why clock-truthful bidding does not constitute an equilibrium: Bidders do not get any surplus from winning three lots, so a better strategy is to submit a headline demand of one unit in the initial round. This would end the auction immediately with both bidders having a positive payoff.

\begin{figure}
    \centering
        \footnotesize{
\begin{tikzpicture}
\draw [->, thick] (0,0)--(11.7,0);
\node [right] at(11.7,0){$\mathbf p$};

\node [thick,blue] at (0.6,1.04){\textbf{Headline demands}};
\node [thick,red] at (0.34,2.1){\textbf{Additional bids}};
\node [thick] at (-.35,-.63){\textbf{Surplus}};

\draw [thick] (0,-0.08)--(0,.08);   %
\node [below] at (0,0) {0}; %
\node [blue,above, rectangle, draw, minimum width = 57pt, minimum height = 15pt] at (0,.2) {$\diamond\diamond\diamond$}; %
\node at (0,1.52) {\textcolor{red}{$\varnothing$}}; %
\node [blue, rounded rectangle, draw, minimum width = 57pt, minimum height = 15pt] at (0,-1.15){$90$};

\node at (1.3,1.52) {$\dots$};
\node at (1.3,.47) {$\dots$};
\node [below] at (1.3,-.15) {$\dots$};
\node at (1.3,-1.15) {$\dots$};

\draw [thick] (2.6,-0.08)--(2.6,.08);   %
\node [below] at (2.6,0) {5}; %
\node [blue,above, rectangle, draw, minimum width = 57pt, minimum height = 15pt] at (2.6,.2) {$\diamond\diamond\diamond$}; %
\node at (2.6,1.52) {\textcolor{red}{$  \varnothing$}}; %
\node [blue, rounded rectangle, draw, minimum width = 57pt, minimum height = 15pt] at (2.6,-1.15){$90-3p$};

\node at (3.9,.5) {$\dots$};
\node at (3.9,1.52) {$\dots$};
\node [below] at (3.9,-.15) {$\dots$};
\node at (3.9,-1.15) {$\dots$};

\draw [thick] (5.2,-0.08)--(5.2,.08);   %
\node [below] at (5.2,0) {10}; %
\node [blue, above, rectangle, draw, minimum width = 57pt, minimum height = 15pt] at (5.2,.2) {$\diamond\diamond\diamond$}; %
\node [red,rectangle, draw, minimum width = 57pt, minimum height = 15pt] at (5.2,1.52) {$3p-30;\,\diamond\,\diamond$};
\node [blue, rounded rectangle, draw, minimum width = 57pt, minimum height = 15pt] at (5.2,-1.15){$60$};
\node [red, rounded rectangle, draw, minimum width = 57pt, minimum height = 15pt] at (5.2,-1.79){$60$};

\node at (6.5,1.52) {$\dots$};
\node at (6.5,.47) {$\dots$};
\node [below] at (6.5,-.15) {$\dots$};
\node at (6.5,-1.47) {$\dots$};

\draw [thick] (7.8,-0.08)--(7.8,.08);   %
\node [below] at (7.8,0) {15}; %
\node [blue, above, rectangle, draw, minimum width = 57pt, minimum height = 15pt] at (7.8,.2) {$\diamond\diamond\diamond$}; %
\node [red,rectangle, draw, minimum width = 57pt, minimum height = 15pt] at (7.8,1.52) {$3p-30;\,\diamond\,\diamond$};
\node [blue, rounded rectangle, draw, minimum width = 57pt, minimum height = 15pt] at (7.8,-1.15){$90-3p$};
\node [red, rounded rectangle, draw, minimum width = 57pt, minimum height = 15pt] at (7.8,-1.79){$90-3p$};

\node at (9.1,1.52) {$\dots$};
\node at (9.1,.47) {$\dots$};
\node [below] at (9.1,-.15) {$\dots$};
\node at (9.1,-1.47) {$\dots$};

\draw [thick] (10.4,-0.08)--(10.4,.08);   %
\node [below] at (10.4,0) {20}; %
\node [blue, above, rectangle, draw, minimum width = 57pt, minimum height = 15pt] at (10.4,.2) {$\diamond\diamond\diamond$}; %
\node [red,rectangle, draw, minimum width = 57pt, minimum height = 15pt] at (10.4,1.2) {$3p-30;\,\diamond\,\diamond$};
\node [red,rectangle, draw, minimum width = 57pt, minimum height = 15pt] at (10.4,1.84) {};
\node [red] at (10.285,1.84) {$3p-60;\,\diamond$};
\node [blue, rounded rectangle, draw, minimum width = 57pt, minimum height = 15pt] at (10.4,-1.15){$30$};
\node [red, rounded rectangle, draw, minimum width = 57pt, minimum height = 15pt] at (10.4,-1.79){$30$};
\node [red, rounded rectangle, draw, minimum width = 57pt, minimum height = 15pt] at (10.4,-2.4){$30$};

\end{tikzpicture}
}
    \caption{Headline demands and additional bids under CMRA-truthful bidding strategy in the illustrative example.}
    \begin{minipage}{.9\linewidth}
\footnotesize{\emph{Notes:} Above the price axis: Diamonds indicate the number of units in the bid; headline demands are at the clock price $p$ and additional bids are at stated prices. Below the price axis: Surplus from each individual bid. The additional bids are such that the bidder is indifferent between winning with the headline demand and the additional bid.}
\end{minipage}
    \label{fig:example}
\end{figure}
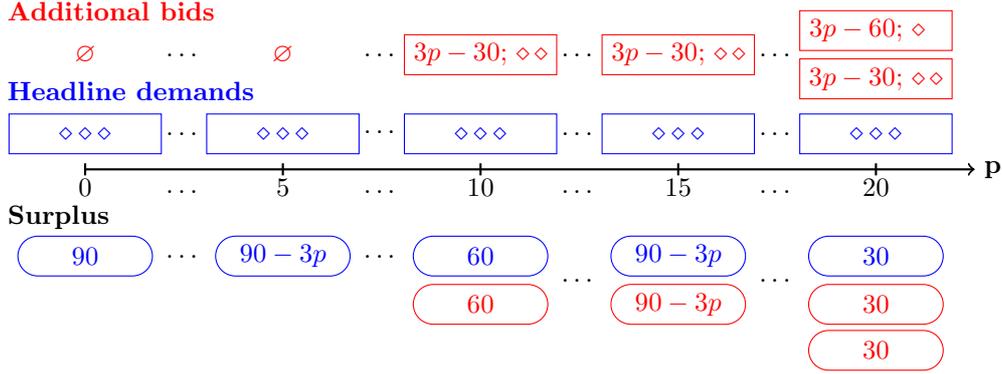

Now consider the presence of additional bids in the CMRA (headline demands continue to be three units at prices below \$30). Figure~\ref{fig:example} illustrates the \emph{CMRA-truthful} strategy in our example. At any clock price up to \$10, the bidder only submits a headline demand of three units. At the clock price of \$10 the bidder is indifferent between winning three lots at linear prices (with a surplus $3 \times \$20 = \$60$) and winning two lots at a price of zero. Hence, the bidder submits an additional bid for a total of \$0 for two lots (note that the price in this additional bid is below the clock price). Indeed, if both bidders submit such bids, there exists an allocation of all four lots to both bidders such that exactly one bid is accepted from each bidder, but it is not revenue maximizing (i.e., it raises \$0 compared to the allocation of three units to one bidder which raises \$30). Hence the auction continues. As the clock price increases, the bidder changes the additional bid for two lots. For example, at a clock price of \$12, the bidder submits an additional bid of \$6 for two lots. At a clock price of \$20, the bidders place an additional bid of \$30 for two lots and add an additional bid of \$0 for one lot; the surplus from each bid is \$30. The auction now ends because there are now revenue-maximizing allocations in which exactly one bid of each bidder is accepted: either the bidders' additional bids on two lots are accepted or one bidder's headline demand of three lots and one bidder's additional bid for one lot are accepted. In any case, the final allocation is efficient but revenue is lower (\$60 vs. \$90) than in the clock auction.
We show that, perhaps surprisingly, CMRA-truthful strategies constitute an ex-post equilibrium in proxy strategies whenever bidders' marginal values are constant (as in this example) or increasing and the caps are symmetric. However, the CMRA-truthful equilibrium does not survive \emph{any} asymmetry in bidders' caps.

There turns out to be another equilibrium in which the bidders can achieve risk-free collusion. In the first round of the auction, the bidders follow the CMRA-truthful strategy, but also add an additional bid for two lots at a price of \$0. The auction ends immediately and both bidders win two units each at a price of zero. If the auction does not end in the first round, the bidders submit CMRA-truthful bids in all subsequent rounds and the auction will proceed as in case of CMRA-truthful equilibrium. 
Note that this first-round demand reduction carries no risk for the bidders in the sense that bidders do not have to give up ``eligibility" to (potentially) end the auction early. In other words, they do not have to lower their headline demands to implement a collusive outcome. The clock auction, on the other hand, allows demand reduction only in headline demands; hence, lowering demand to two lots in early rounds does not allow the bidder to win more units later.
In Section~\ref{sec:risk free}, we propose an alternative activity rule that prevents bidders from following this collusive strategy.

The model used in the paper is different from the illustrative example in three ways. First, we assume for tractability that the good is divisible. Second, we do not assume that marginal values are constant. Instead, we also allow for non-decreasing and increasing marginal values. Third, we allow for bidders to have asymmetric caps and incomplete information about each others' types.

\section{The Combinatorial Multi-Round Ascending Auction}
\label{sec:model}
We now present the general model of the CMRA used in the non-strategic analysis of Section~\ref{sec:truthful}. The strategic analysis will require additional assumptions which we will state in Section~\ref{sec:strategic}. There is a set $M$ of $m\geq 1$ goods in positive and bounded supply and a set $N$ of $n\geq 2$ bidders. Let $X_i\subseteq \mathbb R_+^m$ be the set of bundles which are feasible for bidder $i\in N$. The goods can be indivisible, in which case $X_i \subseteq \mathbb{Z}^m_+$, or divisible, in which case $X_i$ is compact and convex for each bidder $i\in N$. Feasibility constraints could also incorporate caps which we discuss explicitly in Section~\ref{sec:strategic}. Let $\mathcal X\subseteq\bigtimes_{i\in N} X_i$ denote the set of feasible allocations.

Bidder $i$'s utility from winning bundle $x_i\in X_i$ for transfer $t_i$ is quasilinear and equal to $U_i(x_i)-t_i$, where $U_i\colon X_i\to\mathbb R_+$ is (weakly) monotone and $U_i(0)=0$. Let $x^\star\in\arg\max_{ x\in\mathcal X}\sum_{i\in N}U_i( x_i)$ denote the efficient allocation. 
We assume that the total supply of goods is allocated in the efficient allocation.

We map the Danish spectrum auction rules to our setting. The rules for $m=1$ correspond to those of the 2016 auction for the 1800 MHz band, while the rules for the $m>1$ case follow those of the 2019 auction.

\paragraph{Clock and headline demands.} The core of the CMRA is a clock auction: There is a vector of clock prices $p\in\mathbb{R}_{+}^m$ and in each clock round (indexed by $p$), each bidder $i$ reports a demanded bundle $h_i(p)\in X_i$, which is called the \emph{headline demand}.

\paragraph{Additional bids.} An innovation of the CMRA is that bidders can submit \emph{additional bids} in each clock round. The additional bid for bundle $x_i\in X_i$ when the clock prices are $p$ is the price (willingness-to-pay) $A_i(x_i;p) \in \mathbb{R}_{+}\cup\{-\infty\}$. In contrast to the headline demands, the additional bids do not have to be linear in clock prices. Instead, they must be below clock prices, i.e., $A_i(x_i;p) \le px_i$ and must satisfy the activity rule described below. 
For quantities that do not receive any additional bids, we define the additional bids to be negative, i.e., $A_i(x_i;p) = - \infty$ if bidder $i$ has not submitted an additional bid on bundle $x_i$. The auction rules specify that bids on the empty package must be non-positive. 

In each clock round $p$, the headline demand and the additional bids create a bid function $B_i\colon X_i \times \mathbb R_{+}^m \to \mathbb R_{+}\cup\{-\infty\}$ that maps bundles and the clock prices to bids. The bid function $B_i(\cdot;p)$ is then the collection of highest bids expressed in the course of the auction by bidder $i$ at clock prices $p$. For the headline demand the bid function simply gives us $B_i(h_i(p);p) = p h_i(p)$ while for the additional bids the bid function is defined as $B_i(x_i;p) = \sup_{\tilde p \le p} A_i(x_i; \tilde p)$. 

\paragraph{Closing rule.} The auction ends if there is a feasible revenue-maximizing allocation in which a bid by every bidder is accepted. Formally, the auction ends in clock round $p$ if there is a feasible allocation $x\in\mathcal X$ such that
\begin{enumerate}
    \item $x\in\arg\max_{\tilde x\in\mathcal X}\sum_{i\in N}B_i(\tilde x_i;p) $, and 
    \item $B_i(x_i;p) \ge 0$ for all $i\in N$.
\end{enumerate} 
The second requirement guarantees that bidders win nothing only if they bid (zero) on the empty package. If there is no such allocation, the auction continues by raising at least one of the clock prices $p$. 

It is worth comparing the closing rule in the CMRA to the closing rules in the clock auction and the CCA. The clock auction ends as soon as there is no excess demand. A bidder wins nothing only if she drops her demand to zero. The final allocation need not maximize revenue as it does not take past bids into account (unlike the CMRA). The clock phase in the CCA ends when there is no excess demand. Depending on the specific activity rules and the final clock round, it is possible that bidders do not win anything. The CCA selects the final allocation as in closing rule condition~1 above (i.e., ignoring condition~2), but the bidders' payments are weakly lower than $B_i$ due to VCG-pricing.

\paragraph{Price adjustment.}  
All our results hold for the following general class of price adjustment rules.
The clock price vector $p$ starts at zero. In each round indexed by $p$, the auctioneer (continuously) increases the price of at least one of the goods for which the aggregate headline demand is non-zero until the closing rule ends the auction.

CMRA price adjustment rules used in practice belong to the class of rules specified above.
For example, the 2019 Danish spectrum auction rules prescribe considering bidders who are omitted in the revenue-maximizing allocation. For any such bidder $i\in N$ and all goods $k\in M$ headline-demanded by this bidder, re-run the revenue-maximization problem when assuming that the bidder had instead only headline-demanded $q_k$, where $q_k$ is the $k^\text{th}$ entry of $h_i(p)$. If the bidder is still omitted, then (continuously) increase the price of good $k$.\footnote{To see that the Danish price adjustment rule never increases the clock price of a good that was not headline-demanded, observe that if any omitted bidder's headline demand is replaced by $0\in\mathbb R^m_+$, then this bidder cannot be omitted.} The rationale behind this rule is that $p_k$ would not need to change if the market cleared at this price if the bidder headline-demanded the (smaller) package $(0,\dots,0,q_{k},0,\dots,0)$.\footnote{This price adjustment rule can reveal information about the other bidders' bids if $m>1$. An alternative price adjustment rule would only use excess demand in headline demands as long as there is excess demand. If there is no excess demand, one possibility is to increase all clock prices in all clock rounds if information revelation should be avoided.}

\paragraph{Activity rule.} 
The description of the activity rule for general $m$ is lengthy. As our analysis exclusively uses it in the $m=1$ case of Section~\ref{sec:strategic}, we describe it only for this specific instance; in the non-strategic analysis of Section~\ref{sec:truthful}, we consider truthful bidding which should be allowed by any reasonable activity rule. The activity rule currently used in the CMRA translates the headline demands into constraints on the additional bids. Let $m=1$, $p < p'$ and $h_i(p) > h_i(p')$. Suppose $p$ was the highest price at which $h_i(p)$ was demanded and that for all prices $\tilde{p}>p$ bidder $i$ demanded $h_i(p')$. For $x \in (h_i(p'), h_i(p))$ the ``relative cap" activity rule constrains $B_i(x;\tilde p)$ from above by\footnote{If headline demand is $x$ and lowered continuously at clock price $p$, then the slope of the bids at $x$ can be at most $p$.}
\begin{equation*}
    B_i(x;\tilde p)\le B_i(h_i(p');\tilde p) + p'(x - h_i(p')).
\end{equation*}
If $m>1$, then so-called eligibility points would need to be specified.

\section{Outcomes under non-strategic truthful bidding} 
\label{sec:truthful}

In this section, we investigate outcomes under non-strategic truthful bidding. Our motivation is three-fold. First, truthful strategies allow us to introduce the workings of the CMRA without the need to keep track of incentives. Second, truthful bidding provides a natural benchmark in auctions with many bidders. Lastly, these strategies set the stage for the strategic incentives studied in the next section.

We compare the simple clock auction to the CMRA to isolate the effect of the additional bids. As the CMRA is built on a clock auction, any clock auction behavior is also feasible in the CMRA; bidders only have to ignore the possibility of additional bids.

In the clock auction, the meaning of truthful bidding is unambiguous. Bidders can demand only a single package at a time, and this bundle is utility-maximizing if it is truthful. 
Similarly, the headline demand $h_i(p)$ in the CMRA is \emph{truthful} at current price $p$ if $h_i(p) \in \arg\max_{x} U_i(x)-p x$. We further say that a bidding strategy in the CMRA is \emph{clock-truthful} if the bidder only submits truthful headline demands at all clock prices and submits no additional bids.

The CMRA's additional bids allow richer behavior that may be considered truthful. 
Truthful headline demands at all prices elicits a demand curve that reflects the bidder's marginal values. 
This motivates us to interpret truthful bidding in the CMRA as expressing true marginal values for all shares for which this is possible at given clock prices. 
The following ``truthful'' bidding strategy was described by DotEcon:
\begin{quote}
    \emph{The proposed auction design allows bidders to follow alternative bid strategies that give them more control over their risks. For instance, a bidder may submit many bids at all times with a view to maximise surplus. Making a headline bid for the surplus-maximising package and then for all other packages with a bid amount equal to the value of the package minus the surplus on the headline bid would achieve this. \citep{dotecon2016proposal}}
\end{quote}
In other words, DotEcon suggests that bidders place an additional bid for $x_i$ such that the bidder is indifferent between winning $x_i$ for a payment of $A_i(x_i;p)$ and the headline demand at price vector $p$. We say that a bidding strategy is \emph{CMRA-truthful} if the headline demands are truthful and
\begin{equation}\label{eq:additional}
    A_i(x_i;p) =
    \begin{cases}
        U_i(x_i) - V_i(p) &\text{if } U_i(x_i) - V_i(p)\ge 0\\
        -\infty &\text{otherwise}
    \end{cases}
\end{equation}
for all $x_i\in X_i$, where $V_i(p) = \max_x U_i(x) - p x$. The additional bids are generically non-linear in clock prices $p$.

The CMRA-truthful bids progress as follows in the course of the auction. The headline demand is a large bundle at $p=0$ and eventually decreases to the empty package. There may be no additional bids initially. As the value $V_i(p)$ associated with winning the headline demand decreases, previous additional bids are increased and (low) additional bids on new bundles are placed. Both properties result from $V_i(p)$ decreasing in $p$ due to the envelope theorem. Hence, the bidder bids true marginal values for all packages for which this is possible with non-negative bids but shades the bids by bidding less than value (for sufficiently low $p$).\footnote{The CMRA-truthful strategy can be seen as a family of ``truthful strategies'' \citep{Bernheim-Whinston-Menu-Auction} or ``truncation reports'' \citep{day-milgrom-core-selecting}. A difference between the two strategies is that Bernheim-Whinston's strategy prescribes to bid zero on small packages while the CMRA-truthful strategy does not involve bids on small quantities (which we model as being $-\infty$). The difference implies that the CMRA does not end for any ``profit target'' (shading amount) due to the auction's closing rule.}
The bid function converges to the bidder's utility function $U_i$ as clock prices $p$ increase and equal the utility function as soon as the empty package is headline-demanded. 

\subsection{Efficiency}
Our first result proves that the CMRA ends with the efficient allocation if bidders bid CMRA-truthfully. The result holds under fairly general conditions, since the goods can be divisible or indivisible.
\begin{theorem}\label{prop:CMRA truthful}
    CMRA-truthful bidding leads to the efficient allocation.
\end{theorem}

The result might seem surprising at first because, while it assumes truthful bidding, it does not (i) require any conditions on the structure of the valuations, or (ii) stipulate the specifics of the price adjustment process, or (iii) reduce to a dynamic implementation of the VCG mechanism (see Section~\ref{sec:strategic}).
Additional bids play the key role in guiding the CMRA auction towards an efficient allocation. First, note, that the closing rule ensures that the final allocation is efficient. Therefore, if the allocation is not efficient, the auction must adjust prices and the bidders submit more additional bids. As a result, the auctioneer is gathering more information about bidders' marginal valuations (reflected in the additional bids) over the course of the auction and is therefore eventually able to find the efficient allocation. The examples in Section~\ref{sec:example} and Supplemental Appendix~\ref*{online appendix illustration CMRA truthful} illustrate how an efficient outcome may not be reached with headline demands alone (or via a clock auction), but can be found once the additional bids are truthfully submitted.

There are other dynamic (package) auction formats that are efficient under truthful bidding. While some are crafted for a particular class of preferences \citep{ausubel2002ascending,ausubel2006AER,sun2014efficient,fujishige2025universally}, others work for general quasilinear preferences \citep{ausubel2002ascending,DeVries-Schummer-Vohra-JET-2007}. Some of these formats do not have efficient equilibria under strategic bidding. We will examine strategic properties of the CMRA in Section~\ref{sec:strategic}.

\subsection{Revenue}
We now turn to the revenue properties of the CMRA by comparing revenue under CMRA-truthful bidding to revenue under truthful demand revelation in a competitive equilibrium. A competitive equilibrium is defined in the usual way: a price vector $p^\star\in\mathbb{R}^m_+$ and a feasible allocation $x^\star\in\mathcal{X}$ constitute a competitive equilibrium if (i) $x_i^\star\in \arg\max_{x_i \in X_i} U_i(x_i)-p^\star x_i$ for all $i \in N$ and (ii) aggregate demand $\sum_{i\in N} x_i^\star$ must be less than supply for all goods (with the equilibrium price of any good being strictly positive whenever the supply constraint binds). To be able to meaningfully compare the revenue in a competitive equilibrium to revenue under CMRA-truthful bidding, we need that a competitive equilibrium exists and that the clock prices would converge to the competitive equilibrium under CMRA-truthful bidding.

\begin{proposition}\label{prop:revenue-competitive-equ}
    Suppose that a competitive equilibrium $(x^\star,p^\star)$ exists and let $h_i(p^\star)=x_i^\star$ be the headline demands at clock prices $p^\star$ for all $i\in N$ in the CMRA. For any continuous and monotonically increasing sequence of clock prices converging to $p^\star$, it holds that the CMRA ends at $\tilde p^\star\le p^\star$ under CMRA-truthful bidding, and revenue is lower than in the competitive equilibrium $p^\star$.
\end{proposition}

The CMRA ends at a price $\tilde p^\star$ where for at least one $k\in M$ $\tilde p_k^\star<p^\star_k$ if at least one bidder's preference for $x_i^\star$ is strict at $p^\star$.

Proposition~\ref{prop:revenue-competitive-equ} requires equilibrium existence and monotone convergence of prices. Both can be guaranteed in the case of divisible and indivisible goods for large classes of preferences.  
For example, a competitive equilibrium exists if goods (and units) are (gross) substitutes \citep{kelso1982job,Milgrom-Strulovici} (and is unique if goods are also divisible \citep{Wald-ECMA-1951}). Under the assumptions, the clock auction would lead to the competitive equilibrium under truthful bidding \citep{milgrom2000putting}, so, if the CMRA adopted the clock auction's price adjustment rule (i.e., increasing the prices of goods that are headline over-demanded), then the CMRA revenue would be lower than in the clock auction.
In the case of a single good as analyzed in the next section, there is a single clock price that is increased until the auction ends; the CMRA's clock price necessarily converges to the competitive equilibrium. With concave utility functions, a competitive equilibrium exists even when the units of the good are indivisible \citep{henry}. Hence, the revenue in the CMRA is lower  than in the clock auction. Supplemental Appendix \ref*{online appendix illustration CMRA truthful} illustrates the CMRA under truthful bidding and the revenue comparison with the clock auction (also in the case for non-decreasing marginal values).

\section{Outcomes under strategic bidding}
\label{sec:strategic}

We now turn to the analysis of strategic bidding in the CMRA. We focus on the case of two bidders and one perfectly divisible good for tractability and clarity (our negative results would, of course, carry over to a more general case). We first show that neither form of truthful bidding is an equilibrium with decreasing marginal values. In contrast, when marginal values are non-decreasing, CMRA-truthful (but not clock-truthful) bidding can be an equilibrium with symmetric caps. However, this equilibrium is fragile to small asymmetries in the caps. We then characterize a particularly simple but non-truthful equilibrium. 
Finally, in Section~\ref{sec:risk free}, we illustrate how bidders can risklessly collude in the CMRA. The main takeaway from Sections~\ref{sec:strategic} and~\ref{sec:risk free} is that the CMRA does not always provide incentives for truthful bidding and is instead vulnerable to many forms of strategic manipulation.

\subsection{Further assumptions for the strategic analysis}\label{sec:further}

In the strategic analysis of Sections~\ref{sec:strategic} and~\ref{sec:risk free}, there are two bidders ($N=\{1,2\}$) bidding for a perfectly divisible good in unit-mass supply ($m=1$). There is a publicly known \emph{cap} $\lambda_i\in (1/2,1)$ that constrains bidder $i$ to win at most $\lambda_i$, i.e., $X_i=[0,\lambda_i]$.\footnote{\label{footnote:capacity}An alternative interpretation is that $\lambda_i$ is bidder $i$'s capacity \citep{ausubel2014demand} or satiation point where marginal values are at or below the reserve price. In this case, we would define $U_i(x)$ to be strictly increasing on $[0,\lambda_i]$ and flat on $(\lambda_i,1]$, that is, $U_i(x) = U_i(\lambda_i)$ for $x\in [\lambda_i,1]$.} Without loss of generality, let $\lambda_1\ge\lambda_2$. Spectrum caps were present in all real-world implementations of the CMRA. Note that some auctions feature symmetric caps while other auctions restrict bidders asymmetrically. Hence, both cases are of interest. 

There is a value function $U$ which is twice continuously differentiable and parameterized by private types $\theta_i$, i.e., $U_i(x) = U(x;\theta_i)$, where $\theta_i \in [\underline \theta_i, \overline \theta_i]$ and $0 \leq \underline \theta_i<\overline \theta_i<\infty$. Let $u_i(x)$ denote the marginal value, so $U_i(x)=\int_0^x u_i(y)dy$. Marginal values are strictly positive. We say that bidder $i$ is \emph{stronger} than bidder $j$ if $\theta_i\geq\theta_j$. Let values be strictly increasing in $\theta_i$, i.e., $\partial U(x;\theta_i)/\partial \theta_i > 0$ for all $x\in [0,\lambda_i]$. The marginal values are strictly increasing in $\theta_i$, that is, $\partial U/\partial x \partial \theta_i > 0$.  
We assume that conditional on bidder $j$'s type, the distribution of bidder $i$'s type has full support on $[\underline{\theta}_i,\overline{\theta}_i]$ and do not further restrict the joint type distribution due to our focus on ex-post equilibria. 

\paragraph{Strategies and equilibrium.} A \emph{proxy strategy} $(h_i, (A_i(\cdot;p))_{p\ge 0})$ for bidder $i$ consists of a headline demand $h_i$ and a collection $(A_i(\cdot;p))_{p\geq 0}$ of additional bids indexed by clock prices $p$. Throughout, we use \emph{ex-post equilibrium in proxy strategies} as our solution concept. In such an equilibrium, bidders' strategies are proxy strategies that are best responses to the opponent's strategy and any type distribution. In particular, bidders would not choose an alternative proxy strategy if they knew the opponent's type.\footnote{Note that ex-post equilibria refine Bayes-Nash equilibria even in proxy strategies because bidders can update their belief about the opponent's type as the clock price changes.} Similar to \citet{levin2016properties}, we study equilibria in proxy strategies to focus on outcomes that are specific to the CMRA. 
\citet{levin2016properties} study equilibria in the CCA in proxy strategies and point out that dynamic auctions typically feature equilibria in non-proxy strategies in which certain actions trigger reactions of the other bidders (cf. \cite{ausubel-schwartz}). The restriction to proxy strategies avoids the specification of (updated) beliefs about the other bidder's type, which does not play a major role in our analysis due to our focus on ex-post equilibria.\footnote{Moreover, proxy strategies only specify ``on-path" behavior; as the proxy strategy rules out certain bidding histories, the strategy does not specify behavior after such histories. Specifically, there are sequences of additional bids that put non-trivial constraints on the bidding rules. Specifying behavior in such cases is notationally complex. We avoid these issues by considering proxy strategies.} Whenever we say that an ``equilibrium exists'', we mean that an ex-post equilibrium in proxy strategies exists for every type profile; conversely, if we say that an ``equilibrium does not exist'', we mean that there is a type profile which does not admit such an ex-post equilibrium.

\paragraph{Information.} In each round, bidders learn (i) the current clock prices, (ii) whether their previous headline demand appeared in a revenue-maximizing allocation, (iii) whether they appeared in a revenue-maximizing allocation. Due to our focus on proxy strategies, (ii) and (iii) will play no role to our analysis. Moreover, in the 2021 Danish spectrum auction bidders learned (i)--(iii), while in the 2016 and 2019 auctions they only learned (i). Hence, for 2016 and 2019 auctions the restriction to proxy strategies is without loss (except for what may be learned from the price adjustment rule in 2019).

\subsection{No truthtelling with decreasing marginal values}

We first examine whether the clock-truthful or CMRA-truthful strategies form equilibria when bidders have decreasing marginal values. Bidders are said to have \emph{decreasing marginal values} if there are two bidders, one perfectly divisible good, $u_i'(x) < 0$ for $x\in [0,\lambda_i]$, and $u_i(\lambda_i) < u_j(1-\lambda_i)$ for $i,j=1,2$ and all feasible type profiles. The assumption of decreasing marginal values implies that the efficient allocation is interior, i.e., $1-\lambda_j < x_i^\star < \lambda_i$; the efficient allocation $x^\star$ solves $\max U_1(x_1) + U_2(x_2)$ subject to $x_1 + x_2 \le 1$. The assumptions imply that $x_1^\star = 1-x_2^\star$ and that $u_1(x_1^\star) = u_2(1-x_1^\star)$.\footnote{Corner solutions in which one bidder is much stronger than another merely add to casework without affecting the economic content of our results.}

To study the strategic incentives, we recall the CMRA's pricing rule: winning headline demands are priced at the clock price and winning additional bids are priced as bid. Such pricing features appear in clock auctions and (sealed-bid) discriminatory multi-unit auctions. A common property of equilibria in these auctions is that bidders have an incentive to ``shade" their bids below their value \citep{ausubel2014demand,Bernheim-Whinston-Menu-Auction}. This intuition carries over to the CMRA when bidders have decreasing marginal values.

\begin{remark}\emph{ Clock-truthful bidding is not an equilibrium in the CMRA when bidders have decreasing marginal values just like truthful bidding is not an equilibrium in the clock auction. As bidders pay $p x$ for winning $x$ at final clock price $p$, they have an incentive to report a marginally lower willingness-to-pay to end the auction at a lower price as this saves them money on units they win. Such (marginal) deviations are possible only if the other bidder uses a downward sloping demand curve as is natural with decreasing marginal values.}
\end{remark}

The following proposition states that CMRA-truthful bidding also does not form an equilibrium in the CMRA when bidders have decreasing marginal values.
\begin{proposition}\label{theorem:noeqm}
    Suppose that marginal values are decreasing. 
        There is no equilibrium in which each bidder bids CMRA-truthfully.
\end{proposition}

To understand why CMRA-truthful bidding is not an equilibrium, suppose both bidders bid CMRA-truthfully. Theorem~\ref{prop:CMRA truthful} implies that the auction ends only if both bidders bid on their efficient share $x_i^\star$. As soon as they do so, revenue is at a \emph{local} maximum (due to decreasing marginal values). Let $\tilde p$ denote the lowest price at which both bidders bid on $x_i^\star$. At $\tilde p$, however, the joint bid on $x^\star$ is less than $\max_{i} B_i(\lambda_i;\tilde p)$; hence, $x^\star$ is not revenue-maximizing. By Proposition~\ref{prop:revenue-competitive-equ}, we know that at some clock price $\tilde p^\star<p^\star$, the additional bids are sufficiently high so that the efficient allocation $x^\star$ is \emph{globally} revenue-maximizing. Hence, the auction ends at $\tilde p^\star$ at which 
\begin{equation}\label{eq:freeride}
    B_1(x^\star_1;\tilde p^\star) + B_2(x^\star_2;\tilde p^\star) = \max_{i} B_i(\lambda_i;\tilde p^\star).
\end{equation}

For prices $p\in(\tilde p,\tilde p^\star)$, bidders face a threshold problem\footnote{This is related to what happens in other combinatorial auctions where small bidders need to jointly outbid a larger competitor. In our case, the bidders may wish to coordinate additional bids on small quantities in order to outbid their bids on larger quantities.} and the stronger bidder faces two kinds of deviation incentives captured by Eq.~\eqref{eq:freeride}.
The first incentive, highlighted by the proof, is for the stronger bidder to shade her bids on high quantities. This allows the stronger bidder to remove the threshold problem and close the auction early. In Eq.~\eqref{eq:freeride}, the stronger bidder might wish to decrease $B_i(\lambda_i; \tilde{p}^\star)$. 
This incentive alone means that CMRA-truthful bidding is not an equilibrium.
Moreover, the stronger bidder has an incentive to free-ride on her competitor by shading her bids on the efficient shares even more, i.e., bidding lower than dictated by the truthful additional bids. 

We end the discussion of decreasing marginal values with another observation about payments and revenue.
\begin{remark}\label{prop: decreasing marginal values pay more than VCG}
    Let the caps be symmetric, $\theta_i>\theta_j$, and the marginal values decreasing, and bidders bid CMRA-truthfully. The weaker bidder $j$ pays the VCG price; the stronger bidder $i$ pays more than the VCG price.
\end{remark}
We already know from Proposition~\ref{prop:revenue-competitive-equ} that revenue under CMRA-truthful bidding is lower than in a competitive equilibrium, but Remark~\ref{prop: decreasing marginal values pay more than VCG} tells that the revenue is higher than in a VCG auction.

The observation that in the two-bidder case the payments are not equal to the VCG payments allows us to compare the CMRA to the ``menu auction'' of \citet{Bernheim-Whinston-Menu-Auction}.  
In the pay-as-bid menu auction, bidders can use a ``profit target'' strategy: $B_i(x)=\max\{0, U_i(x)-t_i\}$, where $t_i$ is the profit target and $x\in X_i$. If there are two (potentially asymmetric) bidders, there exists a Nash equilibrium of the menu auction in profit-target strategies in which bidders pay VCG prices  \citep[Corollary 1]{Bernheim-Whinston-Menu-Auction}. In the menu auction, incentive constraints determine equilibrium profit targets. CMRA-truthful bidding forms a family of profit target strategies with the profit target (i.e., the indirect utility $V_i(p)$) decreasing in the clock prices (see Eq.~\eqref{eq:additional}). Thus the CMRA determines the final profit target endogenously through the closing condition on the auctioneer's side (i.e., Eq.~\eqref{eq:freeride}). As a result, the profit targets in the CMRA are additionally affected by the closing condition so the CMRA does not necessarily lead to the VCG payments for both bidders.

\subsection{Fragile truthtelling with non-decreasing marginal values}

Instead of decreasing marginal values (capturing substitutes), bidders might have non-decreasing marginal values (capturing complements). Bidders are said to have \emph{non-decreasing marginal values} if there are two bidders, one perfectly divisible good, and $u_i'(x) \ge 0$ for $x\in[0,\lambda_i]$ for $i=1,2$. The efficient allocation must then lie on the boundary. Recall that we assume that $\lambda_1\ge \lambda_2>\frac{1}{2} $. If $\theta_1\ge \theta_2$, then $(\lambda_1,1-\lambda_1)$ is the efficient allocation. To see this, observe that $U_1(\lambda_1) + U_2(1-\lambda_1) > U_1(1-\lambda_2) + U_2(\lambda_2)$ is equivalent to $\int_{1-\lambda_2}^{\lambda_1} u_1(x)dx > \int_{1-\lambda_1}^{\lambda_2} u_2(x)dx$. Both integrals are over the same mass but the left-hand side integrates over higher quantities. As the marginal values are increasing in $\theta$, the inequality holds. In the case of symmetric caps, a similar argument shows that it is efficient that the stronger bidder wins $\lambda$. Now, if $\theta_2 > \theta_1$, then in principle both allocations $(\lambda_1,1-\lambda_1)$ and $(1-\lambda_2,\lambda_2)$ can be efficient. Intuitively, $(\lambda_1,1-\lambda_1)$ can be efficient if $U_1(\lambda_1)$ is much higher than $U_2(\lambda_2)$ due to the asymmetric caps and increasing marginal values while there may be not much difference between $U_1$ and $U_2$ for $x\le \lambda_2$. 

Unsurprisingly, clock-truthful bidding is also not an equilibrium when the bidders have non-decreasing marginal values. 
\begin{remark}\label{rem:clockeqm}\emph{Clock-truthful bidding is not an equilibrium in the CMRA when bidders have non-decreasing marginal values. 
Under clock-truthful bidding, bidders first headline-demand $\lambda_i$ and then drop demand to zero. The auction ends as soon as the first bidder does so. Hence, under clock-truthful bidding, weak bidders would expect to win nothing, therefore submitting a headline demand of $1-\lambda_j$ at a clock price of $0$ is a profitable deviation.}
\end{remark}

Incentives to deviate are markedly different against a CMRA-truthful opponent compared to the clock-truthful one described in Remark~\ref{rem:clockeqm}. Since marginal values are non-decreasing, each bidder's headline demand is $\lambda_i$ (or 0). As long as no bidder $i$ bids on $1-\lambda_j$, $j\neq i$, the auction continues because the revenue-maximizing allocation involves one bidder winning $\lambda_i$, but this allocation has not received bids from both bidders. This changes as soon as one bidder bids zero on the residual supply. That is, there is a \emph{final price} $p_i^f$ at which bidder $i$ bids 0 on $1-\lambda_j$ as this bidder is indifferent between winning $\lambda_i$ for a payment $ p_i^f\lambda_i$ and winning $1-\lambda_j$ for free. Using the indifference condition $U_i(\lambda_i) - p^f_i\lambda_i = U_i(1-\lambda_j)$, we can write bidder $i$'s final price as 
\begin{equation}\label{eq:pif}
    p^f_i = \frac{U_i(\lambda_i) - U_i(1-\lambda_j)}{\lambda_i}.
\end{equation}
However, with asymmetric caps, the additional bid of zero on $1-\lambda_j$ does not always end the auction (as the allocation $(\lambda_i,1-\lambda_i)$ might not be revenue-maximizing), which distorts the incentives away from CMRA-truthful bidding. 
However, in the case of symmetric caps, the weaker bidder makes a bid of zero on $1-\lambda$ before the stronger bidder and this bid ends the auction. This turns out to be sufficient for CMRA-truthful bidding to form an equilibrium under symmetric caps.

\begin{theorem}\label{theorem:increasing-MV-CMRA-truthful-expost}
    Suppose that marginal values are non-decreasing.
\begin{enumerate}[label=(\roman*)]
    \item Let $\lambda_1=\lambda_2$. There exists an equilibrium in which each bidder bids CMRA-truthfully.
    \item Let $\lambda_1>\lambda_2$. There does not exist an equilibrium in which each bidder bids CMRA-truthfully. 
    \end{enumerate}
\end{theorem}
Focusing on the symmetric-caps case, it may be surprising that CMRA-truthful bidding can ever be an (ex-post) equilibrium: Truthful bidding neither forms an equilibrium in the pure clock auction nor in the discriminatory auction. 
Yet, in the CMRA, a combination of these two auction formats, truthful bidding even forms an \emph{ex-post} equilibrium.\footnote{In Appendix \ref{app: more than two bidders} we show that CMRA-truthful bidding also forms an ex-post equilibrium when there are more than two bidders with non-decreasing marginal values and symmetric caps.} However, the existence of this equilibrium is not robust to asymmetries in the bidders' caps. From a practical perspective, it is worth pointing out that some auctions feature symmetric caps. 
However, if $\lambda_i$ denotes the bidder's capacity (see footnote~\ref{footnote:capacity}), then the symmetry assumption may be more unlikely to hold.

Why is CMRA-truthful bidding an equilibrium when bidders have non-decreasing marginal values and symmetric caps? The truthful headline demand is $\lambda$ as long as the clock price is not prohibitively high.  
Consider any deviating bid $D_i(x)$ for quantity $x$. Such a bid must be sufficiently high as it becomes winning only if it is part of a revenue-maximizing allocation. In particular, the deviating bid $D_i(x)$ must satisfy 
\begin{equation*}
    D_i(x) + B_j(1-x;p) \ge B_j(\lambda;p).
\end{equation*}
Plugging in the truthful bids of bidder $j$ 
gives
\[D_i(x) \ge U_j(\lambda) - U_j(1-x).\] 
Observe that the right-hand side is precisely the VCG price in a VCG auction in which bidders bid truthfully. In such a VCG auction, the strong bidder $i$ wins $\lambda$ and pays $U_j(\lambda) - U_j(1-\lambda)$, and the weak bidder $j$ wins $1-\lambda$ and pays $U_i(\lambda) - U_i(\lambda) = 0$. In the CMRA with truthful bidding and symmetric caps, bidders' payments are also exactly these VCG prices, which implies that bidders cannot profitably deviate. 

Asymmetric caps change the auction outcome under CMRA-truthful bidding in a crucial way: The auction does not necessarily end with the first zero-bid on $1-\lambda_j$. In particular, if the bidder with the higher cap (bidder 1) places the zero-bid on $1-\lambda_2$ first, then the auction does not end as bidder 1's headline demand is still revenue-maximizing but does not include bids from both players:
\[B_1(1-\lj;\pif) + B_2(\lj;\pif) = \pif\lj < \pif\li=B_1(\li;\pif).\]
Hence, the auction continues and drives up prices, which impacts the bidders' incentives. If bidder 1 eventually wins $1-\lj$, the payment is strictly positive and bidder 1 has an incentive to end the auction early by deviating to a headline demand of $1-\lj$ in the first round. On the other hand, if the type profile was such that $\pif<\pjf$ and $(\li,1-\li)$ the final (efficient) allocation, then the auction would end at clock price $\pjf$ under CMRA-truthful bidding. However, observe that bidder 1 then prefers winning $1-\lj$ for free over winning $\li$ for transfer $\pjf\li$:
\[U_1(\li) - \li\pjf = U_1(\li) - \frac{\li}{\lj}\left(U_2(\lj)-U_2(1-\li)\right) < U_1(1-\lj) \Leftrightarrow \pif<\pjf.\]
Bidder 1 can win $1-\lj$ for free by submitting an additional bid of zero for $1-\lj$ at clock price 0. Note that bidder~1 would pay an \emph{inflated} VCG price under CMRA-truthful bidding: $\li\pjf$ equals $\li/\lj$ times the VCG price, where the coefficient $\li/\lj>1$ appears due to the asymmetric caps. These arguments show that CMRA-truthful bidding can form an equilibrium only if $\pjf\le\pif$ for all type profiles; in this case, the auction certainly ends at $\pjf$ with bidder~2 making the zero-bid. Next, we show that this necessary condition is not sufficient.

CMRA-truthful bidding is not bidder~1's best response to bidder~2's CMRA-truthful strategy if $\pjf<\pif$ for all type profiles. In particular, bidder~1 can lower the payment for $\li$ from $\pjf\li$ to $p\li$, where $p<\pjf$. In the proof, we also provide a lower bound on $p$ to make the deviation feasible. The lower payment can be guaranteed by headline-demanding $\lambda_1$ until clock price $p$, not making any additional bids, and dropping demand to 0 at clock price $p$. This freezes the price bidder~1 has to pay for $\li$ to $p\li$. Due to the cap asymmetry and $\pjf<\pif$, the auction does not end at $p$ but continues until bidder~2 places the zero-bid on $1-\li$ at clock price $\pjf$. As bidder~1 can secure $\li$ for a lower payment, CMRA-truthful bidding does not form an equilibrium.

\subsection{A simple non-truthtelling equilibrium} 

We now describe a simple bidding strategy which turns out to be an ex-post equilibrium irrespective of whether the marginal values of both bidders are increasing or decreasing. In the symmetric-caps case, the bidders submit a headline demand of $\lambda$ for as long as it gives them non-negative surplus. There are no additional bids except for a single additional bid of $0$ on $1-\lambda$ at clock price $p_i^f$ (at which bidder $i$ is indifferent between winning $\lambda$ and winning $1-\lambda$ for free). As such, the strategy is ``almost clock-truthful'' when marginal values are non-decreasing. We refer to this strategy as the \emph{\flatbid\ bidding} strategy. Given that both players play this strategy, the stronger bidder $i$ wins $\lambda$ at a linear price of $p_j^f$ and the weaker bidder $j$ wins $1-\lambda$ for free. With non-decreasing marginal values, the auction outcome is as under CMRA-truthful bidding, including the VCG prices. However, we know from Theorem~\ref{theorem:increasing-MV-CMRA-truthful-expost} that bidders have incentives to deviate from this outcome when the caps are asymmetric.

For incentive compatibility in the asymmetric-caps case, the price at which the additional zero-bid is made must only be determined by bidders' types rather than also by their caps. This is achieved by bidders bidding on $\lj$ as their headline demand.\footnote{Indeed, any headline demand $x$ such that $1/2<x\leq \lambda_2$ will sustain an analogous equilibrium. The construction may also apply when caps are non-deterministic. For example, if $\lambda_i$ was drawn randomly from $[\lj,\li]$, the bidders could coordinate on the constant strategies with the possible split of $\lambda_2$ and $1-\lambda_2$.} To this end, define $p^f(\theta)=(U(\lj;\theta)-U(1-\lj;\theta))/\lj$. In the \emph{\flatbid\ }strategy, bidder $i$'s headline demand is
    \begin{equation}\label{eq:constant}
        h_i(p) = 
        \begin{cases}
            \lambda_2 &\text{for } p\le \frac{U_i(\lj)}{\lj} \\
            0 &\text{otherwise;}
        \end{cases}
    \end{equation}
bidder $i$ places a single additional bid of $A_i(1-\lambda_2;p^f(\theta_i))=0$.

\begin{theorem}
    Suppose that either both bidders have decreasing marginal values or both bidders have
    non-decreasing marginal values. 
    \begin{enumerate}[label=(\roman*)]
        \item Let $\li=\lj$. 
        There exists an equilibrium in which each bidder chooses the \flatbid\ strategy.
        \item \label{theorem:constant asymmetric} Let $\li>\lj$. There exists an equilibrium in which each bidder chooses the \flatbid\ strategy if and only if $\overline \theta_1$ and $\underline \theta_2$ are such that
        \begin{equation}\label{eq:condition constant}\tag{$\spadesuit$}
             U(\li; \overline \theta_1) \le U(\lj;\overline\theta_1) + U(1-\lj;\underline \theta_2).
        \end{equation}
    \end{enumerate}
    If the marginal values are decreasing, the final allocation is generically inefficient. If the marginal values are non-decreasing, (a) the outcome is inefficient if $(\li,1-\li)$ is efficient; (b) the outcome is efficient if $(1-\lj,\lj)$ is efficient; and (c) the outcome is as under CMRA-truthful bidding if the caps are symmetric. 
    \label{theorem:constant-ex-post}
\end{theorem}
To confirm that the constant bidding strategy is a best response to itself, let $\theta_i \ge \theta_j$. Then bidder $i$ knows that she wins $\lambda_2$ for a price of $p^f(\theta_j)$. She cannot win $\lambda_2$ at a lower price. Winning $\lambda_2$ at $p^f(\theta_j)$ is better than winning $1-\lambda_2$ for free by construction of $p^f$ and $p^f(\theta_j) \le p^f(\theta_i)$. She could win less than $1-\lambda_2$, but this is not profitable. With asymmetric caps, bidder 1 may want to win $\li$ with the same deviation strategy as the one that proves that CMRA-truthful bidding does not form an equilibrium. Then the auction does not end at $p^f(\theta_2)$, but at bidder 2's drop-out price $p=U_2(\lj)/\lj$.\footnote{There is a version of the \flatbid\ strategy that is always an ex-post equilibrium. It involves bidder 2 to ``overbid'' as a threat to deter bidder 1 seeking to win $\li$. In this strategy, bidder 2 keeps the headline demand at $\lj$ until the clock price reaches $p'= U(\li;\overline \theta_1)/\lj$. Bidder 1 then needs to bid at least $p'\lj=U(\li;\overline\theta_1)$ to win $\li$. Note that bidder 2's threat may not be credible, however, as it may involve bids above the willingness-to-pay. A practical implication is that bidders with tighter caps might want to signal ``strength'' disproportionately more.} Bidder 1's deviating bid (and therefore payment) must be at least $U_2(\lj)$ so that the deviation leads to a revenue-maximizing allocation. The deviation is then profitable only if
\[\underbrace{U_1(\li)-U_2(\lj)}_{\text{deviation payoff}} \ge \underbrace{U_1(\lj)-U_2(\lj)+U_2(1-\lj)}_{\text{payoff from the constant strategy}}.\] Condition~\eqref{eq:condition constant} implies that winning $\li$ is not profitable. Now, if $\theta_i \le \theta_j$, then bidder $i$ knows she wins $1-\lambda_2$ for free. She cannot win $1-\lambda_2$ at a lower price. She does not want to win more as this would be more expensive and therefore not profitable by construction of $p^f$. 

Note that condition~\eqref{eq:condition constant} constrains only the highest type of bidder 1 and the lowest type of bidder 2; the equilibrium exists if the bidder with the higher cap cannot also have a much higher type than the low-cap bidder. Bidder 2 can have arbitrarily high types because bidder 2's cap eliminates the potential deviation of winning more than $\lj$. Observe that condition~\eqref{eq:condition constant} can be satisfied for sufficiently similar caps for any given $\overline\theta_1$ and $\underline\theta_2$. The following examples illustrate the characterization in Theorem~\ref{theorem:constant-ex-post}\ref{theorem:constant asymmetric}.

\begin{example}[Quadratic utility]
    Let $\li>\lj$, the utility function be quadratic, $u_i(x)=\theta_i-x$, and increasing, i.e., $\min\{\underline\theta_1,\underline\theta_2\}\ge 1$. Let $u(\lj;\overline \theta_1) = \overline \theta_1 - \lj < \underline \theta_2-1+\lj = u(1-\lj;\underline\theta_2)$, which is a stronger version of the assumption that the efficient allocation is interior; it implies that bidder 1 winning more than $\lj$ is never efficient. This assumption guarantees that the \flatbid\ strategy forms an ex-post equilibrium. Plugging the quadratic utility functions into condition~\eqref{eq:condition constant} leads to
    \[\overline\theta_1 \le \frac{\underline\theta_2(1-\lj) + \frac{1}2\li^2 -\frac{1}2 +\lj-\lj^2}{\li-\lj}.\]
    The assumption that the efficient share is not too large also places an upper bound on bidder 1's types: $\overline\theta_1<\underline\theta_2 - 1+2\lj.$ Whenever this upper bound is satisfied, condition~\eqref{eq:condition constant} also holds as
    \[\overline\theta_1<\underline\theta_2 - 1+2\lj \le \frac{\underline\theta_2(1-\lj) + \frac{1}2\li^2 -\frac{1}2 +\lj-\lj^2}{\li-\lj}.\]
    To see that the latter inequality holds, multiplying by the denominator and collecting the terms leads to
    \[0 \le (1-\li)\left(\underline\theta_2 -\frac{1}2(1-\li)\right) + (\li-\lj)^2.\]
    The right-hand side is true because $\underline \theta_2\ge 1$ and $0<\li<1$.
    \hspace*{\fill}$\blacktriangleleft$
\end{example}

\begin{example}[Linear utility]\label{ex:linconstant}
    Let $\li>\lj$ and the utility function be linear in $x$, i.e., $u_i(x)=\theta_i$. Condition~\eqref{eq:condition constant} reduces to 
    \[\overline\theta_1\le \underline\theta_2\frac{1-\lj}{\li-\lj}.\]
    If $\li=0.7$ and $\lj=0.6$, the highest type of bidder 1 can be up to four times the lowest type of bidder 2 for the constant bidding equilibrium to be sustained. As the caps tend to be as asymmetric as possible ($\li\to 1 $ and $\lj\to1/2$), the equilibrium exists only if bidder 1's highest type is at most (marginally above) bidder 2's lowest type.
    \hspace*{\fill}$\blacktriangleleft$
\end{example}

The \flatbid\ strategy features ``demand expansion'' when the bidders have decreasing marginal values.\footnote{Bidder 1 always reduces demand relative to truthful headline demands for low clock prices.} This may be surprising because one typically associates clock auction and pay-as-bid auction formats with demand reduction rather than demand expansion. Here, the demand expansion leads to an inefficient allocation in which the strong bidder may win an even greater quantity than in the efficient allocation. If there is such bidder asymmetry in the market, one option for the auction designer would be to introduce caps that are tighter on the stronger bidder than on the weaker one. With non-decreasing marginal values, bidder 1 reduces demand compared to truthful bidding while bidder 2 bids ``almost clock-truthfully.''\footnote{The bidding behavior of bidders with decreasing marginal values can be interpreted as them bidding clock-truthfully under non-decreasing marginal values. \cite{janssen2019clock} describe an equilibrium in the CCA in which bidders with decreasing marginal values also act as if their marginal values are non-decreasing (jumps in the demand function).}

The \flatbid\ strategy also provides an opportunity to raise rivals' costs despite the presence of pay-as-bid pricing in the CMRA.\footnote{It is well-known that VCG pricing can provide ample opportunities to raise rival's cost in the CCA \citep{janssen2016spiteful,levin2016properties,janssen2019clock}.} Suppose bidder $i$ knows that the other bidder is stronger and bids according to the \flatbid\ strategy. A best response for bidder $i$ is to demand $1-\lambda_2$ at clock price 0. However, following the \flatbid\ strategy leads to a higher price for bidder $j$. In particular, bidder $i$ with a lexicographic preference for raising bidder $j$'s payment will follow the \flatbid\ strategy; extending the clock beyond $p^f(\theta_i)$ leads to the risk of winning $\lambda_2$ at clock price $p^f(\theta_j)$.

\section{Collusion via risk-free demand reduction}
\label{sec:risk free}
We now discuss how additional bids can be used to collude via demand reduction in the CMRA and we will do so in the two-bidder model of Section~\ref{sec:strategic}. In many dynamic auctions, bidders have incentives to (tacitly) coordinate on low demands in early rounds to close the auction at low prices. 
However, in a clock auction or a CCA, if a bidder reduces demand, the activity rule (monotonicity constraint) requires demand to stay low. Therefore, (unilateral) demand reduction can be a risky strategy.

In contrast, the CMRA's additional bids can be on small quantities while the headline demand is high. In particular, at low clock prices, bidders can use additional bids on small quantities to find a revenue-maximizing allocation at which a bid from each bidder is accepted. Demand reduction using additional bids is therefore risk-free in the sense that the bidder does not lose the ability to demand large quantities at higher clock prices if the coordination at low clock prices is unsuccessful.\footnote{Note that a dominant strategy is truly `risk-free.' In contrast, our definition of risk-free is not about payoffs, but about the ability to demand large quantities following an unsuccessful attempt to end the auction at low clock prices.} 

Concretely, we propose the following collusive strategy in the CMRA. Bidders follow the \flatbid\ strategy as in Theorem~\ref{theorem:constant-ex-post} but augment it by a single additional bid of 0 for quantity $1/2$ in the initial clock round. If both bidders bid this way, the revenue-maximizing allocation immediately assigns half of the supply to each bidder and the revenue is 0. Note that as the headline demand is $\lambda_2$, the bidder can later still demand large quantities if the auction does not end in the first round. This makes the demand reduction strategy risk-free for bidders. We call this modified strategy the \emph{risk-free demand reduction} (RDR) strategy. Formally, the headline demand of the RDR strategy is as in Equation~\eqref{eq:constant} and there are two additional bids: $A_i(1/2;0)=0$ and $A_i(1-\lj;p^f(\theta_i))=0$. 

The next theorem shows that both bidders following the risk-free demand reduction strategy is an ex-post equilibrium in proxy strategies if the bidders are sufficiently similar to each other.\footnote{The risk-free demand reduction strategies form a Bayes-Nash equilibrium if the other bidder is sufficiently strong in expectation.} The existence of this equilibrium relies on a focal allocation that splits the market and is acceptable to every bidder.\footnote{In our model, $(1/2,1/2)$ is such a focal split. In practice, there might be no such focal allocation or indeed many of them (so bidders would need to use multiple additional bids to reach one).}

\begin{theorem}\label{theorem:collusion}
Suppose that both bidders have either decreasing 
or non-decreasing marginal values.
    \begin{enumerate}[label=(\roman*)]
        \item Let $\li=\lj$. 
        There exists an equilibrium in which each bidder chooses the risk-free demand reduction strategy if and only if 
        \begin{equation}\label{eq:risk free condition}\tag{$\clubsuit$}
            U( \lj;\overline \theta_i )-U(1/2;\overline \theta_i)  \leq   U( \lj;\underline \theta_j )-  U( 1 - \lj;\underline \theta_j )
        \end{equation}
        holds for $i=1,2$ and $j=3-i$.
    \item Let $\li>\lj$. There exists an equilibrium in which each bidder chooses the risk-free demand reduction strategy if conditions~\eqref{eq:condition constant} and~\eqref{eq:risk free condition} hold.
    \end{enumerate}
\end{theorem}

Using the \flatbid\ strategy as the continuation limits the potentially profitable deviations. Since the \flatbid\ strategy profile forms an ex-post (continuation) equilibrium, we essentially only have to check incentive compatibility at clock price 0. In particular, to prove that the additional bid on 1/2 at clock price 0 is incentive compatible, it is sufficient to ask whether a bidder prefers winning 1/2 for free over the outcome when both follow the \flatbid\ strategy. In the latter case, the stronger bidder $i$ wins $\lj$ for transfer $p^f(\theta_j)\lj$ and the weaker bidder $j$ wins $1-\lj<1/2$ for free. Hence, it is apparent that the weaker bidder has no incentive to deviate from the RDR strategy. The stronger bidder also prefers the RDR strategy as
\[U_i(1/2) > U_i(\lj) - p^f(\theta_j)\lj = U_i(\lj) - \lj\frac{U_j(\lj)-U_j(1-\lj)}{\lj}, \]
which is implied by Condition~\eqref{eq:risk free condition}. In the symmetric-caps case, the continuation strategies always form an ex-post equilibrium. However, in the asymmetric-caps case, bidder 1 may want to win $\li$ instead of 1/2 for free. To make such a deviation unprofitable, we also impose condition~\eqref{eq:condition constant}.

Condition~\eqref{eq:risk free condition} holds when the types are sufficiently close to each other.\footnote{The RDR strategy can be modified to accommodate more than two bidders: Bidders submit additional bid $A_i(1/n;0)=0$ at a clock price of 0 and choose a suitable continuation strategy. Provided that bidders are sufficiently similar, the strategy profile will form an ex-post equilibrium.} 
The following examples illustrate when the RDR equilibrium exists.

\begin{example}[Quadratic utility]
    Let $u(x;\theta)=\theta-x$. For simplicity, we consider ex-ante symmetric bidders with $\li=\lj=\lambda$ and $\underline\theta_1=\underline\theta_2=\underline \theta$ and $\overline\theta_1=\overline\theta_2=\overline\theta$. We show that for $\underline\theta>1.25$, condition~\eqref{eq:risk free condition} holds whenever the efficient allocation is interior. Under the symmetry assumptions, the efficient allocation is interior if $\overline\theta < \underline\theta -1+2\lambda.$ Condition~\eqref{eq:risk free condition} reduces to 
    \[\frac{1}2\overline\theta(2\lambda-1) + \frac{1}8\le \underline\theta (2\lambda-1) + \frac{(1-\lambda)^2}2. \]
    Rewriting, we obtain $\overline\theta \le 2\underline\theta -\frac{3}4 + \frac{\lambda}2.$ The efficient allocation being interior implies condition~\eqref{eq:risk free condition} if
    \[\overline\theta\le\underline\theta-1+2\lambda\le 2\underline\theta -\frac{3}4 + \frac{\lambda}2.\]
    The latter inequality reduces to $-1/4+3\lambda/2\le\underline\theta.$ The inequality holds for all $\lambda$ if $\underline\theta>1.25.$
\end{example}

\begin{example}[Linear utility]
Let $U(x;\theta)=\theta x$. Condition~\eqref{eq:risk free condition} boils down to $\underline\theta_j(2\lj-1)\ge\overline\theta_i (2\lj-1)/2$ and holds as long as a bidder's strongest type is at most twice the weakest type of the other bidder.
\end{example}

\subsection{An alternative activity rule to mitigate collusion}

We now take a closer look at which part of the auction rules make the demand reduction strategy risk-free and how the rules might be modified to make collusion riskier. Mechanically, the demand reduction in the CMRA works through the additional bids. But it is the activity rule that makes collusion risk-free. Recall that in the RDR strategy the bidder expresses zero marginal value between 1/2 and $\lambda$ in the initial clock round. If the auction does not end at $p=0$, the bidder raises the bid on $\lambda$ (through the headline demand the bid on $\lambda$ is $p\lambda$) while keeping the additional bid on $1/2$ constant at $0$. Note that the CMRA uses an activity rule in which only headline demands place constraints on the bidding function and only for quantities between two headline demands. But activity rules are often introduced to ensure consistent bidding along the auction price path, usually in the sense of revealed preferences \citep{ausubel2014market,ausubel2020revealed}. However, the activity rule used in the CMRA does not transform additional bids into constraints on the bid function; as a result, observed bids  can violate revealed preference.\footnote{
To see this, note that at $p=0$, the bidder bids $B(1/2;0)=0$ and $B(\lambda;0)=0$. Hence, the bidder is indifferent between $1/2$ and $\lambda$: $U(1/2)=U(\lambda)$. At $p > 0$, the utility-maximizing quantity cannot be $\lambda$ because $U(\lambda) - p\lambda < U(1/2) - p/2$ and $\lambda>1/2$.}

We therefore propose the following \emph{alternative activity rule} that also transforms the additional bids into constraints.\footnote{Our alternative activity rule is deliberately straightforward and we propose it to show clearly how it affects the equilibria we discuss. It is an interesting research direction to consider whether other activity rules, such as the GARP activity rule, would also have good properties in our context \citep{ausubel2014market,ausubel2020revealed}.} Assuming a finite number of bids, let $X(p)$ denote the set of all quantities bidder $i$ submits non-negative bids on at clock price $p$: $X(p) = \{x:B_i(x;p)\ge 0\}.$ Consider any adjacent $x,x'\in X(p)$ with $x<x'$,\footnote{That is, for any $x''\in(x,x')\Rightarrow x''\neq X(p)$.} and denote $B_i(x';p) - B_i(x;p) = z$. If $z\ge0$ and $y\in(x,x']$, then for all $p'\ge p$ the alternative activity rule requires that \[B_i(y;p') \le B_i(x;p') + z\frac{y-x}{x'-x}.\]
The alternative activity rule strictly strengthens the activity rule used in the real-world implementations of the CMRA (described in Section~\ref{sec:model} and used throughout our analysis). 
To see this, suppose that bidder $i$ reduces his headline demand from $x'$ to $x$ at clock price $p$. Then the bidding function is such that $B_i(x';p)-B_i(x;p) = p(x'-x)=z$. The relative cap (with respect to the headline demands) implies $B_i(y;p') \le B_i(x;p') + z(y-x)/(x'-x)$ for $x<y<x'.$

\paragraph{Effect on risk-free demand reduction.} The alternative activity rule renders the RDR strategy infeasible. Suppose behavior in the first clock round is as according to the RDR strategy: headline-demand $\lambda$ at clock price $p=0$ and submit additional bid $A(1/2;0).$ Hence, the only two quantities that receive non-negative bids are $1/2$ and $\lambda$ and $B(\lambda;0)-B(1/2;0)=0$. The alternative activity rule then specifies that for $p>0$ and $y>1/2$ the bidding function must satisfy $B(y;p)\le B(1/2;p).$ This constraint conflicts with the RDR strategy, which features $B(\lambda;p) = p\lambda > B(1/2;p)=0$ for $p$ slightly larger than 0. Moreover, after the initial round behavior of the RDR strategy, the alternative activity rule rules out headline demands above $1/2$ at $p>0$. To see this, suppose the bidder headline-demands $x>1/2$. Then the alternative activity rule requires $B(x;p)=px \le B(1/2;p)$. The auction rules also require $B(1/2;p)\le p/2$, which leads to a contradiction.

\paragraph{Effect on other collusive strategies.} The alternative activity rule does not, of course, rule out all collusive strategies. In particular, the bidders may attempt to collude in headline demands alone---this is similar to the sort of collusion that is familiar from standard clock auctions \citep{brusco-lopomo-restud-2002,GRIMM20031557}. However, such collusion is risky because if the headline demand is reduced and the auction does not end, then the bidder is constrained to bidding on lower quantities for the rest of the auction.

\paragraph{Effect on truthful bidding.} The activity rule does not rule out truthful bidding. At low prices, a truthful bidder submits a bid for the maximum desired quantity so the activity rule---which aims to prevent a bidder from submitting low initial bids---has no bite. Moreover, note that CMRA-truthful bidding continuously increases the interval of quantities that receives non-negative bids. For any interval $[a,\lambda_i]$ such that $B(a;p)\ge 0$, the \emph{marginal} bids on $[a,\lambda_i]$ do not change for $p'>p$. Let us revisit the example in Section~\ref{sec:example}. Each bidder headline-demands 3 units and submits an additional bid of 0 on 2 units at clock price of 10. The corresponding bids are $B(3;10) = 30$ and $B(2;10)=0$, which leads to the constraint $B(3;p) \le B(2;p) + 30$ for $p>10$. Note, however, that the additional bid for two units is always $B(2;p)= 3p-30=B(3;p)-30.$ 
Clock-truthful bidding is also not affected by the alternative activity rule.

\paragraph{Effect on the constant strategy.} The alternative activity rule requires the \flatbid\ strategy to be modified for $p>p_i^f.$ One natural modification is to maintain the indifference between the headline demand $\lambda$ and $1-\lambda$ by increasing the additional bid on $1-\lambda$ to $B_i(1-\lambda;p) = U_i(1-\lambda) - U_i(\lambda)+ p\lambda.$ We now show that the modified strategy is feasible. The activity rules requires
\[B_i(\lambda;p)\le B(1-\lambda;p) + p_i^f\lambda.\]
Plugging in $B_i(\lambda;p)=p\lambda$ and the (truthful) bid for $1-\lambda$ reduces the rule to $p_i^f\le p_i^f$. Hence, the profile of \flatbid\ strategies can be modified to form an ex-post equilibrium under the alternative activity rule.

While the alternative activity rule can mitigate collusion without a substantive impact on other equilibrium strategies, it might affect bidding in ways that our model does not take into account. 
In particular, bidders need to anticipate the effect of the extra bidding constraints on the ability to submit bids later in the auction and therefore to keep track of (potentially) many bundles as prices evolve. As a result, bidders might be more likely to ``miss'' bids and find it more difficult to manage their budgets.

\section{What theory predicts about bidding in practice}
\label{sec:realworld}

In the previous sections, we have identified four patterns of bidding: clock-truthful, CMRA-truthful, the \flatbid\ strategy, and risk-free demand reduction. 
Some of these form equilibria, others do not. 
In equilibrium there can be demand expansion, demand reduction, and (CMRA-) truthful bidding. We now discuss how theory and auction data can be combined to learn something about bidding behavior.

Figure~\ref{fig:dynamics} summarizes which bidding patterns can be explained by the strategies we have described and suggests which bidding patterns in the CMRA would require new theoretical results. While the bidding strategies are inspired by our equilibrium analysis in Sections~\ref{sec:strategic} and~\ref{sec:risk free}, our predictions do not rely on equilibrium bidding.

\begin{figure}
\footnotesize{
\begin{tabular}{cc|ccc}
 & \multicolumn{1}{c}{} & \multicolumn{3}{c}{\textbf{Additional bids in bidding data}}\tabularnewline
\multirow{4}{*}{\begin{turn}{90}
\textbf{Winning bids}
\end{turn}} &  & Few early, more later & Many early & None or few\tabularnewline
\cline{2-5} \cline{3-5} \cline{4-5} \cline{5-5} 
 & Headline only & NA & NA & Clock-truthful\tabularnewline
 & Headline/additional & CMRA-truthful (non-dec.) & NA & \Flatbid\ bidding\tabularnewline
 & Additional only & CMRA-truthful (dec.) & RDR & NA\tabularnewline
\end{tabular}
}
\bigskip
\caption{Predicted bidding patterns under different bidding strategies and marginal valuations.  \label{fig:dynamics}}
\begin{minipage}{.9\linewidth}
\footnotesize{\emph{Note:} ``NA'' are bidding patterns that cannot be explained by our results. RDR stands for risk-free demand reduction.}
\end{minipage}
\end{figure}

In our model, each bidding pattern we have analyzed has different implications for how the auction ends. If only headline demands are winning, the pattern of bidding is consistent with clock-truthful bidding (or with other clock auction behavior such as standard demand reduction) and we can rule out CMRA-truthful bidding, \flatbid\ bidding and the risk-free demand reduction strategy. If a mix of headline and additional bids is winning, the pattern of bidding is consistent with CMRA-truthful bids and \flatbid\ bidding and we can rule out clock-truthful bidding and the risk-free demand reduction strategy. Finally, if only additional bids are winning, the pattern is consistent with the risk-free demand reduction strategy and CMRA-truthful bidding under decreasing marginal values, but we can rule out CMRA-truthful bidding in the case of non-decreasing marginal values, as well as clock-truthful bidding and \flatbid\ bidding. Of course, we cannot distinguish between CMRA-truthful bidding under non-decreasing marginal values and \flatbid\ bidding from winning data only because they are outcome equivalent.

Under CMRA-truthful bidding, a bidder starts with none or few additional bids. In the course of the auction, they not only add new additional bids but also raise existing additional bids. 
Under the risk-free demand reduction strategy, a bidder starts with a headline demand of a large quantity and additional bids for smaller quantities.\footnote{In our model, bidders only submit one additional bid in the risk-free collusive strategy, but in practice with many lots across many bands, it might make sense to submit a number of additional bids which would reflect salient ways of splitting the market. Indeed, bidders have many opportunities to coordinate on a collusive market split at clock prices higher than the reserve.} This cannot happen under truthful bidding because a bidder cannot be indifferent between a large and small quantity of spectrum at equally low prices. Under risk-free demand reduction, we therefore expect high variance in the bid quantities in early rounds and for the auction to end quickly.
Under the \flatbid\ bidding strategy, the headline demand is constant and additional bids only come late in the auction. The additional bid in this case is low but is made at a relatively high clock price.

While the ideal test of bidding behavior would examine actual bidding data, most telecom regulators are reluctant to publish bidding data in spectrum auctions because it can contain highly sensitive commercial information. Therefore, published auction data is often quite restrictive and do not allow us to unambiguously identify bidding strategies. In the next section we examine Danish CMRAs for which the DEA only publishes aggregated payments and allocations for each bidder. As we will show, despite the coarse data, our theoretical predictions can still be used to assess whether observed outcomes are consistent with the described patterns.

\section{The CMRA in Danish spectrum auctions}
\label{sec:Danish auctions}

We now discuss the outcomes of three Danish spectrum auctions that used the CMRA.\footnote{We do not consider the 2020 Norwegian spectrum CMRA. As this auction allocated licenses in an unusually large number of different frequency bands, it is probable that the main task of the auction was achieving coordination among the bidders rather than creating competition for scarce licenses. Excess supply in the auction is evidence for this.} In each auction there were three bidders whom we refer to as A, B, and C for clarity.\footnote{The bidders were Hi3G Denmark ApS, TDC Net A/S, and
TT-Network P/S, respectively.} Each auction allocated lots with and without a coverage obligation. The structure of the auction was such that first the lots with a coverage obligation were allocated (at reserve prices), and then a CMRA was used for the remaining lots. The CMRA first allocated band-specific generic lots. A subsequent assignment phase allocated the specific frequencies. All auctions featured symmetric spectrum caps. 

While we do not have access to detailed bidding data, we are able to make plausible inferences about possible bidding dynamics from allocation and payment data alone. Our approach is to check whether winning bidders' payments were linear or non-linear in quantities, which indicates whether headline or additional bids were winning. Riskless demand reduction is unlikely if headline bids are winning, bidders win their capped quantities, or final prices are much higher than reserve. 

The outcomes changed as bidders became more acquainted with the auction format. In the first auction, it appears that bidders did not use additional bids to end the auction. In the second auction, one bidder basically won the spectrum cap and there is evidence that additional bids were winning. A bidder may have ended the auction with a low additional bid that was made on a small package. In our theoretical framework, this would amount to the \flatbid\ strategy or to CMRA-truthful bidding with non-decreasing marginal values. In the third auction, there is again evidence that additional bids were winning. The prevalence of additional bids in later auctions suggests that bidders might have become more familiar with the auction format over time.

\subsection{Denmark's 1800 MHz spectrum auction}

In 2016, the DEA sold licenses in the 1800 MHz band. The bidders were bidding for 2x65 MHz paired frequencies in the 1800 MHz frequency band. The auction first allocated three 2x10 MHz blocks with a coverage obligation non-competitively: Each bidder won a 2x10 MHz block at the reserve price of DKK 50 million.\footnote{On 30 September 2016, DKK 1 = \$0.15.} The remaining 2x35 MHz were sold in a CMRA. After assigning the blocks with the coverage obligation, there were spectrum caps that allowed each bidder to win at most 2x20 MHz in the CMRA (which corresponds to a $\lambda$ of 0.57).

The CMRA allocated seven lots (2x5 MHz blocks). Each bidder was allowed to win at most four blocks in the CMRA. The reserve price was DKK 25 million. The CMRA allocated generic blocks; specific frequencies were allocated in an assignment stage after the CMRA. The assignment stage used Vickrey-nearest pricing \citep{day2012quadratic}.

Bidder A won 2x20 MHz for a total price of DKK 300,159,486. Bidder B won 2x20 MHz for a total price of DKK 300,159,486. Bidder C won 2x25 MHz for a total of DKK 425,239,229. Each bidder won 2x10 MHz with the coverage obligation, so bidders A and B both won two lots in the CMRA. Bidder C won three lots in the CMRA. 

It is likely that only headline demands were winning. The first evidence is that bidders A and B paid exactly the same amounts. Moreover, after subtracting the reserve price for the 2x10 MHz with the coverage obligation, the payment per lot in the CMRA was exactly DKK 125,079,743 \emph{for each bidder}. We view this as evidence that the final clock price was DKK 125,079,743. 

The CMRA ending with headline demand only is only consistent with clock-truthful bidding. While it is probable that bidders did not submit truthful headline demands but reduced their headline demands, there is no evidence that they used additional bids to end the auction as they would in CMRA-truthful bidding, in the \flatbid\ strategy, or in risk-free demand reduction.

\subsection{Denmark's 700 MHz, 900 MHz and 2300 MHz auction}

In early 2019, the DEA auctioned licenses for frequencies in the 700 MHz, 900 MHz, and 2.3 GHz bands. The DEA used the CMRA to sell 16 blocks across two bands. Revenue was almost DKK 2.2 billion. A 40 MHz block in the 2.3 GHz band was unsold. We believe that it did not sell due to an associated coverage obligation.

There was a large difference in bidders' payments. In Supplemental Appendix~\ref*{app:2019Danish}, we argue it is plausible that bidder B won their headline demand, bidder C won with an additional bid, and bidder A either won their headline demand or with an additional bid. Bidder B won the maximum quantity allowed by the spectrum cap (excluding the unsold lot with the coverage obligation that received no bids), which is consistent in our model with CMRA-truthful bidding under non-decreasing marginal values or with the \flatbid\ strategy. Bidder C won a small quantity at a relatively low price, which is also consistent with these two strategies.

\subsection{Denmark's 2021 auction}

In 2021, the DEA allocated frequencies in the 1.5 GHz, 2.1 GHz, 2.3 GHz, 3.5 GHz and 26 GHz bands. Again, the DEA first auctioned off blocks with coverage obligations and then used the CMRA to sell the rest. The DEA used the CMRA to auction a total of 39 blocks across all the five bands. There were three bidders and revenue was just above DKK 2 billion. The auction ended with no lots unsold. If all lots had sold at reserve price, revenue would have been DKK 865 million. In Supplemental Appendix~\ref*{app:2021Danish}, we argue that at least some additional bids were winning as linear closing prices seem unlikely. Again, this suggests that bidders were not simply using headline demands.

\section{Conclusion}
\label{sec:conclusion}
In this paper, we provided the first theoretical analysis of the CMRA, an auction format that has been used in several European spectrum sales. We find that the auction is efficient whenever bidders bid truthfully. Our analysis of strategic bidding suggests that truthtelling equilibria are vulnerable to bidder asymmetries and that inefficient equilibria might emerge. Crucially, however, the CMRA is prone to risk-free collusive bidding when bidders are sufficiently symmetric. This paper therefore underlines the importance of rigorous theoretical analysis of auction formats in the process of auction design.

Our findings suggest settings in which the CMRA might be a viable auction format if collusion concerns are robustly allayed by an appropriate activity rule. First, the CMRA is more likely to have efficient equilibria in the presence of complements. Second, settings with symmetric caps are less prone to deviations from truthtelling. Third, bidder asymmetry helps to mitigate risk-free collusion incentives. Overall, if the designer wishes to sell complements to asymmetric bidders with symmetric caps, then the CMRA appears to be a practical design alternative.

There are several avenues for further work. First, one could develop the theoretical analysis by looking at other equilibria in our model or by extending the model to heterogeneous goods. Second, one could take our theoretical predictions to real-world or experimental bidding data. Third, one could analyze bidder incentives under richer utility functions that include, for example, spitefulness or risk aversion. Finally, one could explore other alternative activity rules and the associated trade-offs in this auction format.

\appendix
\setcounter{theorem}{0}
\renewcommand{\thetheorem}{\Alph{section}.\arabic{theorem}}
\renewcommand{\thelemma}{\Alph{section}.\arabic{lemma}}

\section{Omitted Proofs}\label{app:proofs}

\subsection{Proofs of Section~\ref{sec:truthful}}

\begin{proof}[Proof of Theorem~\ref{prop:CMRA truthful}]
    Let $\mathcal X_i(p)$ denote the set of bundles bidder $i$ bids on at clock price $p$:
    \[\mathcal X_i(p)= \{x_i\in X_i: B_i(x_i;p)\ge 0\}.\]
    Let $\mathcal X (p) = \mathcal X \cap \bigtimes_{i \in N}\mathcal X_i (p) $ denote the set of feasible allocations that receive non-negative bids from each bidder. Recall that the CMRA ends if 
    there is a revenue-maximizing allocation in which (exactly) one bid by every bidder is accepted. Hence, the auction ends at price vector $p\in\mathbb R^m_+$ if there exists $x \in \mathcal X(p)$ such that $\sum_{i\in N} B_i(x_i;p)\ge \sum_{i\in I} B_i(y_i;p)$ for all $y \in \mathcal X$ and $I\subseteq N$.
    
    Let bidders bid CMRA-truthfully. If the efficient allocation receives bids from all the players (i.e., $x^\star \in\mathcal X(p)$), it must eventually be implemented because it maximizes the sum of bids on $\mathcal X(p)$:
    \[x^\star \in\arg\max_{x\in\mathcal X(p)} \sum_{i\in N} B_i(x_i;p) = \arg\max_{x\in\mathcal X(p)} \sum_{i\in N} U_i(x_i) - V_i(p);\]
    note that for the allocations that receive bids from all bidders, the bid function is utility minus a constant. The constants do not affect the winner determination problem. 
    There might be another allocation that receives higher bids (that does not receive bids from all bidders). However, since $V_i(p)$ decreases in $p$, the efficient allocation is revenue-maximizing for sufficiently high $p$.

    We now show that the CMRA cannot end at clock prices $p$ for which the efficient allocation is not in $\mathcal X(p)$. First, if $x^\star \notin\mathcal X(p)$, then $\mathcal X(p)$ could be empty, in which case the auction would continue.
    
    Let $x\in\mathcal X(p)$ be the allocation that maximizes revenue on $\mathcal X(p).$ Let $I\subseteq N$ be the set of
    bidders who bid non-negatively on $x_i^\star$ at clock price $p$, i.e., $i\in I\Leftrightarrow B_i(x_i^\star;p)\ge 0$. 

    We now show that $I$ is non-empty if $\mathcal X(p)\neq \emptyset$. First, $x\in\mathcal X(p)$ means that $B_i(x_i;p) =U_i(x_i)-V_i(p)\ge 0$ for all $i$, implying that $\sum_{i\in N} U_i(x_i)-V_i(p) \ge 0$. If $I$ were empty, then $U_i(x_i^\star)-V_i(p) <0$ for all $i$ so that $\sum_{i\in N} U_i(x_i^\star)-V_i(p)<0$. However, the two inequalities contradict efficiency. 
    
    At clock price $p$, all $n$ bidders bid positively on $x_i$ (by the definition of $\mathcal X(p)$). It follows that
    \begin{align*}\sum_{i\in N} B_i(x_i;p) &= \sum_{i\in N} U_i(x_i) -V_i(p)\le \sum_{i \in N} U_i(x_i^\star)-V_i(p) \\&\le \sum_{i \in N}\max\{ U_i(x_i^\star) - V_i(p),0\} =\sum_{i\in I} B_i(x_i^\star;p),
    \end{align*}
    where the first inequality uses the efficiency of $x^\star$ and the second inequality uses $U_i(x_i^\star) - V_i(p) < 0$ for $i\notin I$. 
    Therefore, $x$ is not (overall) revenue-maximizing and the auction continues. Above we showed that $I$ is non-empty.
\end{proof}

\begin{proof}[Proof of Proposition \ref{prop:revenue-competitive-equ}]    
    We first show that the CMRA ends at a clock price $\tilde p^\star$ along the price trajectory converging to $p^\star$. To prove that the CMRA ends at $\tilde p^\star \le p^\star$, suppose CMRA-truthful bidding has led to clock price $p^\star$. Let $y\in \mathcal X$ and $I\subseteq N$. It follows that 
    \begin{equation}\sum_{i\in I} B_i(y_i;p^\star) = \sum_{i\in I} U_i(y_i)-U_i(x_i^\star) + p^\star x_i^\star \le\sum_{i\in I}p^\star  y_i \le \sum_{i \in N} p^\star x_i^\star = \sum_{i\in N}B_i(x_i^\star;p^\star),\label{eq:proof prop 1} 
    \end{equation}
    where the first inequality follows from $x_i^\star$ being each bidder's headline demand. The second inequality follows from  the total supply of goods being allocated in the efficient allocation. Hence, the efficient allocation receives bids from all bidders at $p^\star$ and is revenue-maximizing. The CMRA must end at $p^\star$ at the latest. However, it can also end at $\tilde p^\star\le p^\star$.
    
    The CMRA ends with at least one clock price of $\tilde p^\star$ strictly below the corresponding in $p^\star$ if at least one bidder's preference for $x_i^\star$ is strict at $p^\star$: the first inequality in Equation~\eqref{eq:proof prop 1} is then strict. Consequently, by continuity of the price trajectory, the auctioneer's auction-ending indifference condition holds at clock prices $\tilde p^\star \le p^\star$ with one inequality being strict.
    
    CMRA-truthful bidding leads to ex-post revenue 
    \[\sum_{i\in N} B_i(x_i^\star;\tilde p^\star) = \sum_{i\in N} U_i(x_i^\star) - U_i(h_i(\tilde p^\star)) + \tilde p^\star h_i(\tilde p^\star) \le \sum_{i\in N} \tilde p^\star x_i^\star \le\sum_{i\in N} p^\star x_i^\star, \]
    where the first inequality follows from each bidder's truthful headline demand and the second from $\tilde p^\star \le p^\star.$
\end{proof}

\subsection{Proofs of Section~\ref{sec:strategic}}

\begin{proof}[Proof of Proposition \ref{theorem:noeqm}]
    Consider a type profile with distinct types and suppose both bidders bid CMRA-truthfully. The auction ends at price $\tilde p^\star$ such that 
    \begin{equation*}
        B_1(x_1^\star;\tilde p^\star) + B_2(x_2^\star;\tilde p^\star) = \max_i B_i(\lambda_i;\tilde p^\star).
    \end{equation*}
    Without loss, let bidder $i$ denote the bidder with the higher bid on $\lambda_i$ at clock prices $\tilde p^\star$: $B_i(\lambda_i;\tilde p^\star )\ge B_j(\lambda_j;\tilde p^\star)$. If bidder $i$ wants to win $x_i^\star$, bidder $i$ has an incentive to keep $B_i(\lambda_i;p )$ weakly below  $B_j(\lambda_j;p)$ so that the auction ends at lower clock prices (and at a lower bid $B_i(x_i^\star;p)$). If bidder $i$ does not want to win $x_i^\star$, then this can obviously be done. In both cases, bidder $i$ deviates from CMRA-truthful bidding.
\end{proof}

\begin{proof}[Proof of Remark~\ref{prop: decreasing marginal values pay more than VCG}]
To prove the claim, let $\theta_i>\theta_j$, which implies $B_i(\lambda;p)\ge B_j(\lambda;p)$. The proof of Proposition~\ref{prop:revenue-competitive-equ} shows that the CMRA ends at clock price $\tilde p^\star$ such that $B_i\left(x_i^\star; \tilde p^\star\right) + B_j\left(x_j^\star;\tilde p^\star\right) = B_i(\lambda; \tilde p^\star)$, which is equivalent to
\begin{align}
U_i\left(x_i^\star\right) - V_i(\tilde p^\star) + U_j\left(x_j^\star\right) - V_j(\tilde p^\star) = U_i(\lambda) - V_i(\tilde p^\star).
\end{align}
It follows that weak bidder $j$ pays the VCG price: $B_j(x_j^\star;\tilde p^\star)=U_j(x^\star_j) - V_j(\tilde p^\star) = U_i(\lambda) - U_i(x_i^\star)$. 

For the strong bidder $i$'s payment, observe that $B_i\left(x_i^\star;  p\right) + B_j\left(x_j^\star; p\right) = B_j(\lambda;  p)$ holds for $p<\tilde p^\star$. Consequently, bidder $i$ would pay the VCG price at this price, i.e., $B_i(x^\star_i;p)=U_j(\lambda)-U_j(x_j^\star)$. However, $B_i(x_i^\star;p)$ increases in $p$ so that bidder $i$ bids (and pays) more than the VCG price at clock price $\tilde p^\star$.
\end{proof}

\begin{proof}[Proof of Theorem \ref{theorem:increasing-MV-CMRA-truthful-expost}]
    Let $\li=\lj=\lambda$. Suppose bidder $j$ bids CMRA-truthfully. If bidder $i$ also bids in such a way and $\theta_i\ge \theta_j$, then the auction ends with bidder $i$ winning $\lambda$ for a payment of $p_j^f\lambda$. If $\theta_i<\theta_j$, then bidder $i$ wins $1-\lambda$ for free. In the first case the surplus is \(U_i(\lambda) - U_j(\lambda) + U_j(1-\lambda).\label{eq:surplus_flatbid}\) In the second case the surplus is $U_i(1-\lambda)$.
    
    Suppose bidder $i$ wants to win $x$ for payment $D_i(x)$. As bidder $j$'s headline demand is $\lambda$, bidder $i$'s bid $D_i(x)$ at price $p$ is winning only if 
    \begin{equation*}
        D_i(x) + B_j(1-x;p) \ge B_j(\lambda;p). 
    \end{equation*}
    This inequality is binding in the optimal deviation as otherwise the bid $D_i(x)$ can be decreased. Plugging in the expressions for bidder $j$'s bids yields
    \begin{equation*}
        D_i(x) = U_j(\lambda) - U_j(1-x).
    \end{equation*}
    
    Consider $\theta_i\ge \theta_j$. The deviation is profitable only if the surplus from winning $x$ is better than winning $\lambda$, i.e.,
    \begin{equation*}
        U_i(x) - D_i(x) \ge U_i(\lambda) - p_j^f \lambda.
    \end{equation*}
    Plugging in the expression for $D_i(x)$ gives
    \begin{equation*}
        U_i(x) - U_j(\lambda) + U_j(1-x) \ge U_i(\lambda) - U_j(\lambda) + U_j(1-\lambda).
    \end{equation*}
    Note that this is a (weak) contradiction as it is efficient that bidder $i$ wins $\lambda$ (strict if $x\neq \lambda$). We conclude that bidder $i$ does not have an incentive to deviate.
    
    Now consider $\theta_i<\theta_j$. The deviation is profitable if
    \begin{equation*}
        U_i(x) - U_j(\lambda) + U_j(1-x) \ge U_i(1-\lambda),
    \end{equation*}
    where the left-hand side is the expected utility of winning $x$ for a payment $D_i(x)$ and the right-hand side is the expected utility from following the CMRA-truthful strategy. The only $x$ for which the inequality holds is $x=1-\lambda$ as bidder $j$ winning $\lambda$ is efficient. Hence, there is no profitable deviation.

    Let the caps be asymmetric. We first prove two lemmas.
    \begin{lemma}\label{lemma:pif}
        Let marginal values be non-decreasing and let $\li>\lj$. CMRA-truthful bidding forms an ex-post equilibrium in proxy strategies only if $\pjf\le\pif$ for all feasible type profiles. 
    \end{lemma}
    \begin{proof}[Proof of Lemma~\ref{lemma:pif}]
        First, suppose $\pif<\pjf$ and $(\li,1-\li)$ is efficient. We show that the auction ends at clock price $\pjf$ with the efficient allocation. To see this, bidder 1 bids zero on $1-\lj$ at clock price $\pif$ so the allocation $(1-\lj,\lj)$ receives bids from both players. The allocation is not revenue-maximizing since $\li>\lj$; the CMRA continues. The price at which the auction may end with allocation $(1-\lj,\lj)$ is such that
            \[U_1(1-\lj) - U_1(\li)+p\li + p\lj = p\li.\]
        The price that solves the equality is
        \[p=\frac{U_1(\li) - U_1(1-\lj)}{\lj},\]
        which is larger than $\pif$. Moreover, as $(\li,1-\li)$ is efficient, this price is larger than $\pjf$. Consequently, bidder 2 bids 0 on $1-\li$ before the joint bid on allocation $(1-\lj,\lj)$ exceeds $p\li$. The auction ends at $\pjf$ with the efficient allocation. However, bidder 1 prefers winning $1-\lj$ for free over $\li$ for transfer $\pjf\li$:
        \[U_1(1-\lj) -0> U_1(\li)-\pjf\li,\]
        which is equivalent to $\pif<\pjf.$ 
        
        Second, consider type profiles so that $\pif<\pjf$ and $(1-\lj,\lj)$ is efficient. We show that the clock must end at price $p=(U_1(\li)-U_1(1-\lj))/\lj$ with the efficient allocation, and bidder 1 pays $p(1-\lj)$. To show this, as $(1-\lj,\lj)$ is efficient, the prices satisfy
        \[\pif=\pifull < \frac{U_1(\li) - U_1(1-\lj)}{\lj} < \pjfull =\pjf.\]
        At $\pif$, the efficient allocation receives bids from both bidders; however, the joint bid on it is not revenue-maximizing for prices slightly above $\pif$. This changes at $p=\frac{U_1(\li) - U_1(1-\lj)}{\lj}$. The auction does not end earlier with allocation $(\li,1-\li)$ as bidder 2 does not bid on $1-\li$. It follows that bidder 1 could have won the same share for free by submitting an additional bid of zero for $1-\lj$ at clock price 0. In both cases, bidder 1 has a profitable deviation, so CMRA-truthful bidding forms an equilibrium only if $\pjf\le\pif$.
    \end{proof}

    \begin{lemma}\label{lemma:profitable deviation}
        Let marginal values be non-decreasing and let $\li>\lj$. Let the bidders' types be such that $\pjf<\pif$. If bidder 2 bids CMRA-truthfully, then bidder 1 has no incentive to bid CMRA-truthfully.
    \end{lemma}

    \begin{proof}[Proof of Lemma~\ref{lemma:profitable deviation}]
        If both bidders bid CMRA-truthfully, then the auction ends at clock price $\pjf$ with the allocation $(\li,\lj)$. Bidder 1's total payment is $\pjf\li.$ This is because at this clock price this is the first allocation that receives bids from both bidders and the allocation is revenue-maximizing.

        Bidder 1 has the following profitable deviation: Headline-demand $\li$ until the clock price reaches $p\in ( (U_2(\lj)-U_2(1-\li))/\lambda_2,\pjf)$, at which point the headline demand is dropped to 0. 
        No further bids are made by bidder 1. Hence, $B_1(\li;p') = p\li$ for $p'\ge p.$

        If bidder 1 plays this strategy, the auction does not end at clock price $p$, because there is no revenue-maximizing allocations $(\li,y)$, where $0\le y \le 1-\li$, that has received non-negative bids from both bidders; the auction continues until $\pjf$.

        The deviation is profitable only if the auction does indeed lead to allocation $(\li,1-\li)$ at clock price $\pjf$; allocation $(\li,1-\li)$ must be revenue-maximizing:
        \[\underbrace{B_1(0;\pjf)}_{=0} + \underbrace{B_2(\lj;\pjf)}_{\pjf\lj} < \underbrace{B_1(\li;\pjf)}_{=p\li} + \underbrace{B_2(1-\li;\pjf)}_{=0}. \]
        Therefore, for $p$ such that $\pjf>p > \frac{U_2(\lj)-U_2(1-\li)}{\li}$, the deviation does not change the final allocation compared to truthful bidding but reduces bidder 1's total payment from $\pjf\li$ to $p\li$.
    \end{proof}
    Due to marginal utilities being strictly increasing in $\theta$, there is no type space (with at least three distinct types) so that $\pif=\pjf$ for all type profiles. To see this, note that $p_i^f$ only depends on $i$'s type. Consider a type profile such that $\pif=\pjf$. Changing $\theta_i$ while keeping $\theta_j$ constant leads to $\pif<\pjf$ or $\pif>\pjf$ . In the first case, Lemma~\ref{lemma:pif} implies that truthful bidding does not form an equilibrium. In the second case, Lemma~\ref{lemma:profitable deviation} implies that bidder 1 can profitably deviate from truthful bidding.
\end{proof}

\begin{proof}[Proof of Theorem \ref{theorem:constant-ex-post}]

The first observation is that $\theta_i\ge\theta_j\Leftrightarrow p^f(\theta_i)\ge p^f(\theta_j)$. To see this, note that $p^f(\theta)=(U(\lj;\theta)-U(1-\lj;\theta))/\lambda_2 = \int_{1-\lj}^{\lj}u(x;\theta)dx/\lj$. The assumption $du/d\theta>0$ implies that $p^f$ increases in $\theta$. Moreover, $p^f(\theta)<U(\lj;\theta)/\lj$ for all feasible $\theta$.

Now suppose that both bidders follow the \flatbid\ strategy with headline demand in Equation~\eqref{eq:constant}. Bidder $i$ wins $\lj$ for transfer $p^f(\theta_j)\lj$ if  $\theta_i\ge\theta_j$ and wins $1-\lj$ for free if  $\theta_i<\theta_j$. The proof follows similar steps as the proof of Theorem~\ref{theorem:increasing-MV-CMRA-truthful-expost}. 

Suppose that instead of following the \flatbid\ strategy, bidder $i$ wants to win $x_i$ with bid $D_i(x_i)$ at clock price $p$. As bidder $j$'s headline demand is $\lambda_2$ for any clock price below $U_j(\lj)/\lj$, bidder $j$'s bid on $\lambda_2$ is $B_j(\lambda_2;p)=\min\{p,U_j(\lj)/\lj\}\lambda_2$. To be part of the revenue-maximizing allocation, the bid $D_i(x_i)$ must satisfy 
\[
    D_i(x_i) + B_j(x_j;p) \ge B_j(\lambda_2;p),
\]
where $x_i + x_j\le 1$. 

Consider $p < p^f(\theta_j)$. As bidder $j$'s headline demand is $\lambda_2$ and bidder $j$ places no additional bids, bidder $i$ faces
\[
    B_j(x_j;p) =
    \begin{cases}
    -\infty &\text{for } 0\le x_j < \lambda_2\\
    p\lambda_2 &\text{for } x =\lambda_2 .
    \end{cases}
\]
Hence, the only feasible $x_i$ is $x_i\le 1-\lambda_2$. Any such $x_i$ can be won for free. The best $x_i$ is $1-\lj$. If $\theta_i < \theta_j$, then bidder $i$ cannot do better by deviating from the \flatbid\ strategy. If $\theta_i\ge \theta_j$, then bidder $i$ prefers the \flatbid\ strategy as this leads to a higher expected utility:
\[U_i(\lj)-p^f(\theta_j)\lj\ge U_i(1-\lj) \Leftrightarrow p^f(\theta_i)\ge p^f(\theta_j).\]

Consider $p \in [p^f(\theta_j),U_j(\lj)/\lj)$. At clock price $p^f(\theta_j)$, bidder $j$ adds the additional bid on $1-\lambda_2$, which changes the bid function at $1-\lambda_2$ to $B_j(1-\lambda_2;p) = 0$. Now the revenue-maximizing allocation can be $(x_i,1-\lambda_2)$ for $x_i\le\lj$ if $D_i(x_i) \ge B_j(\lambda_2;p)$. In the best case, the inequality is binding. Hence, if $D_i(x_i)$ is winning, then bidder $i$'s utility is at most $U_i(x_i) - B_j(\lambda_2;p)$. The $x_i$ that maximizes the utility is $x_i=\lambda_2$. If $\theta_i \ge \theta_j$, then bidder $i$ wins $\lambda_2$ for $p^f(\theta_j)\lambda_2$, which is at least as good as winning $\lambda_2$ for price $B_j(\lambda_2;p) = p\lambda_2 \ge p_j^f\lambda_2$. Hence, there is no profitable deviation when bidder $i$ is stronger. If bidder $i$ is weaker, then the utility of winning $\lambda_2$ for $B_j(\lambda_2;p)$ is less than winning $1-\lambda_2$ for free, which can be achieved by the \flatbid\ strategy. Hence, there is also no profitable deviation if bidder $i$ is weaker than bidder $j$. 

Finally, consider $p \ge U_j(\lj)/\lj$. Bidder $j$ bids 
\[
    B_j(x_j;p) =
    \begin{cases}
    0 &\text{for } x_j \in \{0,1-\lambda_2\}\\
    U_j(\lj) &\text{for } x =\lambda_2 \\
    -\infty &\text{else. }
    \end{cases}
\]
Winning $x_i\le \lj$ now becomes more expensive compared to the previous case. Bidder 1 might also win $x\in(\lj,\li]$ at clock price $p$. The lowest (and only) clock price at which this is possible is $p=U_2(\lj)/\lj$. At this price, bidder 2 bids 0 on 0 so that an allocation in which bidder 1 wins more than $\lj$ can involve bids by both bidders. The deviating bid must be sufficiently high so that it is revenue-maximizing:
\[D_1(x) + B_2(0;p)\ge B_2(\lj;p)=U_2(\lj).\]
It follows that bidder 1 can win $x$ for payment $U_2(\lj).$ A strategy that leads to this outcome is to headline-demand $x$ until clock price $\tilde p = U_2(\lj)/x$ and then stop bidding. Note that this deviation is the same as the one that proves that CMRA-truthful bidding is not an equilibrium with non-decreasing marginal values (Theorem \ref{theorem:increasing-MV-CMRA-truthful-expost}). Due to positive marginal values and a payment that is independent of the quantity, the optimal quantity is $x=\li$.

Bidder 1's deviation payoff is then $U_1(\li) - U_2(\lj)$. The deviation is profitable if 
\[U_1(\li) - U_2(\lj)\ge U_1(\lj) - U_2(\lj)+U_2(1-\lj).\]
Due to marginal values being increasing in $\theta$, if the highest type of bidder 1, $\overline \theta_1$, has no incentive to deviate against the lowest type of bidder 2, $\underline\theta_2$, then no type of bidder 1 has an incentive to deviate. It follows that condition~\eqref{eq:condition constant} is sufficient for the profile of \flatbid\ strategies to form an ex-post equilibrium. 

Moreover, condition~\eqref{eq:condition constant} is also necessary because otherwise there are type profiles so that bidder 1 has an incentive to deviate along the lines discussed above.

With decreasing marginal values, the \flatbid\ strategy profile leads to allocation $(\lj,1-\lj)$, which is not interior and hence cannot be efficient by assumption for all type profiles.

If marginal values are non-decreasing and $(\li,1-\li)$ is efficient, then the equilibrium outcome is inefficient. If $(1-\lj,\lj)$ is efficient, then $\int_{1-\li}^{\lj}u_2(x)dx \ge \int_{1-\lj}^{\li}u_1(x)dx$. As both integrals are over the same mass, but over lower $x$ for bidder 2, bidder 2 must have a higher $\theta$. Hence, the profile of \flatbid\ strategies leads to the efficient outcome. With symmetric caps and with non-decreasing marginal values, the outcome is identical to the outcome under CMRA-truthful bidding and to truthful bidding in a VCG auction.
\end{proof}

\subsection{Proofs of Section~\ref{sec:risk free}}
\begin{proof}[Proof of Theorem~\ref{theorem:collusion}]
If bidder $i$ does not want to win $1/2$ at clock price 0, then the profile of continuation strategies forms an ex-post equilibrium in proxy strategies by Theorem~\ref{theorem:constant-ex-post}. Hence, the only question is whether bidder $i$ prefers winning 1/2 for free over the continuation payoff of the \flatbid\ strategy profile. 

Let $\theta_i\ge\theta_j$. Then bidder $i$ wins $\lj$ for transfer $U_j(\lj)-U_j(1-\lj)$ by deviating from winning $1/2$ for free in the initial clock round. The deviation payoff is worse as:
\begin{align*}
    U_i(\lj) - U_i(1/2) &\le U(\lj;\overline\theta_i) - U(1/2;\overline\theta_i) \\
    &\le U(\lj;\underline\theta_j) - U(1-\lj;\underline\theta_j)\le U_j(\lj)-U_j(1-\lj).\\
\end{align*}
The first and the last inequalities use that marginal values are strictly increasing in type. The second inequality uses condition~\eqref{eq:risk free condition}.

Conversely, if condition~\eqref{eq:risk free condition} does not hold, then the highest type facing the lowest competitor's type strictly prefers to win $\lj$ for transfer $p^f(\underline\theta_j)\lj$ over winning $1/2$ for free.

Let $\theta_i<\theta_j$. Sticking to the RDR strategy gives bidder $i$ a quantity of 1/2 for free. The deviation of not implementing the 50-50 split leads to a payoff of $1-\lj$ for free. There is no incentive to deviate.
\end{proof}

\section{More than two bidders}
\label{app: more than two bidders}

We now analyze the robustness of the previous results with respect to the number of bidders $n$. We start by looking at CMRA-truthful bidding when $m=1$, the good is perfectly divisible, and $U$ has non-decreasing marginal values. We focus on this case because truthful bidding is not an equilibrium with decreasing marginal values (Proposition~\ref{theorem:noeqm}). Moreover, we restrict attention to symmetric caps because we know that CMRA-truthful bidding is not an equilibrium with asymmetric caps, even in the two-bidder case (Theorem~\ref{theorem:increasing-MV-CMRA-truthful-expost}). Let $1/n<\lambda <1$. 

With non-decreasing marginal values, there are $n^\star := \lceil1/\lambda\rceil$ winners in the efficient allocation ($\lceil x \rceil$ denotes the smallest integer that is larger than $x$). Wlog, let us rank the bidders according to type so that bidder 1 has the highest and bidder $n$ has the lowest type. Then bidder $i$ with $i<n^\star$ wins $\lambda$ and bidder $n^\star$ wins $\hat x:=1-(n^\star-1)\lambda$; $\hat x$ is equal to $\lambda$ if $\lambda=1/n^\star$. Bidder $i>n^\star$ does not win anything in the efficient allocation.

Theorem~\ref{prop:CMRA truthful} implies that auction ends with the efficient allocation when all bidders bid CMRA-truthfully. Let $p_i^f(x)$ denote the clock price at which bidder $i$ bids zero on $x$ under CMRA-truthful bidding:\[p_i^f(x) = \frac{U_i(\lambda)-U_i(x)}{\lambda}.\]
Let $p^F$ denote the clock price at which the auction ends.

When all bidders are winners in the efficient allocation (i.e., if $n=n^\star$ or if $1/n<\lambda< 1/(n-1)$), then the clock ends at $p^F=p_{n^\star}^f(\hat x)$. This case is analogous to the two-bidder case studied above. 
When $n>n^\star$, then the auction does not end at $p_{n^\star}^f(\hat x)$. At $p_{n^\star}^f(\hat x)$, bidder $n^\star$ bids 0 on $\hat x$ but bidder $n^\star+1$ bids something strictly positive on $\hat x$. Hence, the efficient allocation is not revenue-maximizing and the revenue-maximizing allocation does not include bids by all bidders; the auction continues. At clock price $p_{n^\star+1}^f(0)=U_{n^\star+1}(\lambda)/\lambda$, bidder $n^\star+1$ stops bidding. However, bidder $n^\star+1$'s bid on $\hat x$ is still higher than bidder $n^\star$'s until the clock price is 
\[p^F=\frac{U_{n^\star+1}(\hat x) - U_{n^\star}(\hat x) +U_{n^\star}(\lambda)}{\lambda},\]
which is such that $B_{n^\star}(\hat x;p^F) = U_{n^\star+1}(\hat x).$
Note that this price satisfies $p^F>p_{n^\star}^f(\hat x)$ and $p^F\le p_{n^\star}^f(0)$. Moreover, $p^F>p_{n^\star+1}^f(0)$ because marginal values are increasing in $\theta$. Importantly, bidders pay VCG prices. Bidder $n^\star$ bids $U_{n^\star+1}(\hat x)$ on $\hat x$, which is the externality $n^\star$ imposes on $n^\star+1$ (and the other losing bidders). Bidder $i\in \{1,2,\dots,n^\star-1\}$ pays $p^F\lambda= U_{n^\star+1}(\hat x) - U_{n^\star}(\hat x) + U_{n^\star}(\lambda)$ (which is the externality of bidder $i$ having a higher type so that bidder $n^\star$ does not win $\lambda$ and bidder $n^\star+1$ does not win $\hat x$). 

\begin{theorem}
    Let there be one perfectly divisible good and $n\ge 2$ bidders with symmetric caps $\lambda\in (1/n,1)$ and non-decreasing marginal values. CMRA-truthful bidding forms an efficient ex-post equilibrium.
\end{theorem}
\begin{proof}
    We have already argued that CMRA-truthful bidding leads to the efficient allocation. We now prove that no bidder has an incentive to deviate from truthful bidding even if they know the other bidders' types.
    
    \textbf{Case 1.} Let $n=n^\star$. As we argued above, under CMRA-truthful bidding, the auction stops at $p^F=\min_i p_i^f(\hat x) = p_{n^\star}^f(\hat x).$ 
    
    Suppose bidder $i$ wants to win $x_i$ at $p\le p^F$. If $x_i\le \hat x$, then this quantity can be won for free but 
    \[U_i(x_i)\le U_i(\hat x) \le U_i(\lambda)-p\lambda\]
    since $p\le p_i^f(\hat x)$. Note that share $\lambda$ can only be won at $p_{n^\star}^f(\hat x)$ (or at higher prices if $i=n^\star$). If $\hat x < x_i < \lambda$, then the revenue-maximizing allocation can only be such that bidder $i$ wins $x_i$, $n-2$ bidders win $\lambda$ and bidder $n^\star$ wins $1-x_i-(n-2)\lambda>\hat x$ because low types bid higher than high types (except that all bidders bid $p\lambda$ for $\lambda$ as long as $p\le p^f_n(0)$). Hence, bidder $i$ needs to bid at least $D_i(x_i)$ to win $x_i$ in clock round $p$, where $D_i(x)$ must satisfy
    \[D_i(x_i) + B_{n^\star}(1-x_i-(n-2)\lambda) + (n-2)\lambda\ge (n-1)p\lambda.\]
    The right-hand side contains the revenue when all bidders except for bidder $i$ win $\lambda$, which is a feasible allocation but not one in which one bid by every bidder is accepted. The inequality reduces to $D_i(x_i)\ge U_{n^\star}(\lambda) - U_{n^\star}(1-x_i-(n-2)\lambda).$ The optimal deviation is such that bidder $i$ pays the VCG price, i.e., the externality imposed on bidder $n^\star$. Expected utility is then $U_i(x_i) - U_{n^\star}(\lambda) + U_{n^\star}(1-x_i-(n-2)\lambda)$, which is convex in $x_i$ and maximized by $x_i=\lambda$.    

    Suppose bidder $i$ wants to win $x_i$ at $p> p^F$. Buying less than $\hat x$ is not profitable because this quantity can be obtained for free but it is not preferred to the outcome of CMRA-truthful bidding by construction of the strategy. Buying $\lambda$ only becomes increasingly costly. Winning $x_i\in (\hat x,\lambda)$ is possible even at $p\le p^F$ but not profitable. 
    
    \textbf{Case 2.} Let $n > n^\star$. Note that a deviation by bidder $i$ changes the outcome of at most two other bidders; their identify depends on the deviating bidder's. 
    
    Consider first bidder $i=n^\star$. If bidder $i$ wants to win less than $\hat x$, then bidder $n^\star+1$ wins more; the other bidders are not affected. Suppose bidder $i$ wants to win $x_i$ less by bidding $D_i(\hat x - x_i;p)$ in clock round $p$. This deviating bid is part of a revenue-maximizing allocation only if
    \[D_i(\hat x - x_i;p) + B_{n^\star+1}(x_i) \ge B_{n^\star+1}(\hat x).\]
    Bidder $n^\star+1$ bidding truthfully implies that the deviating bid must be at least $U_{n^\star+1}(\hat x) - U_{n^\star+1}(x_i).$ Since this is the VCG price, bidder $i$ has no incentive to deviate from truthful bidding, i.e., $0\in\arg\max_{x_i}U_i(\hat x-x_i) -U_{n^\star+1}(\hat x)+ U_{n^\star+1}(x_i)$. Suppose bidder $i=n^\star$ wants to win $x_i$ more, so $\hat x+x_i$ in total. This can only be done when bidder $n^\star-1$ wins $\lambda-x_i$. Then the deviating bid is part of a revenue-maximizing allocation only if
    \[D_i(\hat x + x_i;p) + B_{n^\star-1}(\lambda - x_i;p) \ge B_{n^\star-1}(\lambda;p).\]
    The deviating bid must be at least the VCG price $U_{n^\star-1}(\lambda) - U_{n^\star-1}(\lambda-x_i)$; it is not profitable to win more. Note that it cannot be the case that bidder $i$ wins so much more that also bidder $n^\star-2$ is affected. If bidder $n^\star$ wins $x_i$ more that $\hat x$, than the cap requires $\hat x+x_i\le \lambda$. Hence, even if bidder $n^\star$ wins $\lambda$, bidder $n^\star-1$ wins $\hat x$ and bidder $n^\star-2$ wins $\lambda$.

    Consider bidder $i>n^\star$, that is, a bidder who does not win anything if bidding CMRA-truthfully. Such a bidder $i$ can only win more than in the efficient allocation by deviating. Depending on how much bidder $i$ wants to win, this reduces how much bidder $n^\star$ wins or it can reduce what bidders $n^\star$ and $n^\star-1$ win. In the first case, the deviating bid must be at least $U_{n^\star}(\hat x) - U_{n^\star}(\hat x - x_i)$, where $x_i$ is the amount bidder $i$ wants to win. Since this is the VCG price, there is no profitable deviation; the optimal $x_i$ is $0.$ In the second case, bidder $n^\star$ does not win anything after the deviation so that the deviating bid must be at least $U_{n^\star-1}(\lambda) - U_{n^\star-1}(\lambda - x_i) + U_{n^\star} (\hat x)$; the externality imposed on the other bidders. None of the two possible deviations can be profitable.

    Consider bidder $i<n^\star$. Bidder $i$ can only win less than in the efficient allocation due to the cap. Depending on how much bidder $i$ wants to win less, there are two possible cases. First, it can be that only bidder $n^\star$ wins more, and, second, that bidders $n^\star$ and $n^\star+1$ win more. In the first case, bidder $i$ has to bid at least $U_{n^\star}(\lambda) - U_{n^\star}(\hat x+x_i)$, where $x_i$ is the amount bidder $i$ wants to win less. Maximizing expected utility $U_i(\lambda - x_i) - U_{n^\star}(\lambda) + U_{n^\star}(\hat x+x_i)$ with respect to $x_i$ leads to $x_i=0$ due to non-decreasing marginal values. Second, suppose bidder $i$ wishes to win so much less that bidder $n^\star+1$ becomes a winner. The deviating bid must then be such that 
    \[D_i(\lambda-x_i;p) + B_{n^\star}(\lambda;p) + B_{n^\star+1}(\hat x -\lambda+ x_i;p)\ge B_{n^\star}(\lambda;p) + B_{n^\star+1}(\hat x;p).\]
    The deviating bid must therefore be at least $U_{n^\star+1}(\hat x) - U_{n^\star+1}(\hat x -\lambda + x_i)$. It is not profitable for bidder $i$ to deviate.
\end{proof}

Finally, we check whether the constant strategy forms an equilibrium. We have already noted that the $n=n^\star$ case is essentially as in in the two-bidder case. Hence, the profile of constant strategies forms an ex-post equilibrium if $n=n^\star$ for decreasing and non-decreasing marginal values (with symmetric caps). However, if $n>n^\star,$ then bidder $n^\star+1$ has a profitable deviation. Instead of submitting a single bid on $\hat x$ with value 0, it is optimal to bid $\epsilon$ on $\hat x$. Doing so guarantees that bidder $n^\star+1$ rather than bidder $n^\star$ wins $\hat x$. Hence, the profile of constant strategies does not form an ex-post equilibrium if $n>n^\star$.

\singlespacing
\bibliographystyle{chicago}
\bibliography{bib}

\newpage
\onehalfspacing
\pagenumbering{roman}
\renewcommand{\thesection}{SA.\arabic{section}}
\renewcommand{\thetable}{SA.\arabic{table}}
\renewcommand{\thefigure}{SA.\arabic{figure}}
\renewcommand{\theproposition}{SA.\arabic{figure}}
\setcounter{section}{0}
\setcounter{table}{0}
\setcounter{figure}{0}
\setcounter{proposition}{0}
\setcounter{footnote}{0}

\setcounter{section}{0}
\begin{center}
    \huge\textsc{{Supplemental Appendix}}\\
    \LARGE{The Combinatorial Multi-Round Ascending Auction}\\
    \large{Bernhard Kasberger and Alexander Teytelboym}\\
    \large{\today}
    
\end{center}

\section{Details of the 2019 Danish auction}\label{app:2019Danish}

Table \ref{tab:2019} summarizes the supply in the 2019 auction and the auction outcome. The licenses in the 900 MHz band were not allocated through a CMRA. In the 900 MHz band, there were 2x30 MHz paired frequencies available. These licenses came with a coverage obligation. We call the lots in the 900 MHz band A lots.

The supply in the CMRA was six 2x5 MHz blocks (B lots) of paired frequencies in the 700 MHz band, four 5 MHz blocks of unpaired frequencies in the 700 MHz band (D lots), one block 40 MHz block in the 2.3 GHz band with a coverage obligation (E lot), and six 10 MHz blocks in the 2.3 GHz band (F lots). 

There was no reserve price on the lots with a coverage obligation. The reserve price per B lot was DKK 95 million. The reserve price per D lot was DKK 25 million. The reserve price per F lot was DKK 25 million.

The following spectrum caps were in place. Each bidder was allowed to win at most one block in the 900 MHz band. Across the paired blocks in the 700 MHz and 900 MHz bands, each bidder was allowed to win at most four lots. Bidders were not allowed to win more than 60 MHz in the 2.3 GHz band. There was no restriction on the number of blocks a bidder could win in the unpaired 700 MHz band.

The auction outcome was as follows. Bidder A paid DKK 485.2 million for one A lot and two B lots. Bidder B paid DKK 1620 million for one A lot, three B lots, four D lots, and six F lots. Bidder C paid DKK 107.6 million for one A lot and one B lot. Hence, the 40 MHz lot in the 2.3 GHz spectrum with the coverage obligation was unsold. Note that bidder B received the maximum quantity permitted by the spectrum caps, which is consistent with the CMRA-truthful and \flatbid\ strategies. For this reason, it is likely that bidder B won with their headline demand.

We now examine whether the bidders paid linear prices. Recall this would suggest that only headline demands were winning. As the A lots were traded at a reserve price of 0, bidder C paid DKK 107.6 million for a single B lot. Bidder A paid DKK 485.2 million for two B lots. As bidder A paid more than four times bidder C's payment, we take this as evidence that at least one additional bid was winning. In particular, we speculate that bidder C won with an additional bid. Winning a small package at low cost is consistent with the \flatbid\ strategy and with CMRA-truthful bidding. 

While we think it is likely that bidder B won with a headline demand, it is not clear whether bidder A won with a headline or an additional bid. Bidder B paid DKK 1135 million more than bidder A for also winning another B lot, the four D lots, and the six F lots. If bidder B won the headline demand, then $1135 = p_B + 4 p_D + 6 p_F$. Suppose bidder A won the headline demand, implying $p_B = 485.2/2=242.6$ and $892.4 = 4p_D + 6p_F$. Moreover, assume that $p_D = p_F$ as the D and F lots have the same reserve price. Linear prices then imply that $p_D = p_F = 89.24$, which is not implausible. Conversely, if bidder B won their headline demand and final prices for D and F were, say, about 65, then this would imply $p_B = 485$. In particular, bidder A would have bought two B lots with an additional bid at half price. Hence, we cannot rule out the possibilities that bidder A won their headline demand or with an additional bid. Finally, we do not see any signs that the auctioned ended as in the risk-free demand reduction equilibrium.

\begin{table}
    
    \centering
    \footnotesize{
    \begin{tabular}{llrrrrr}
    \toprule
    \toprule
    Lot     & Description                   & Supply & R  & Bidder A  & Bidder B  & Bidder C\\
    \midrule
    \multicolumn{7}{l}{\textbf{CMRA}}\\
    B   & 2x5 MHz in the 700 MHz band   & 6     & 95     &  2        &  3        & 1  \\
    D   & 5 MHz in the 700 MHz band    & 4     & 25     &           &  4        &   \\
    E   & 40 MHz in the 2.3 GHz band & 1 & 0 &        &  \\
    F   & 10 MHz in the 2.3 GHz band    & 6     & 25     &           &  6        &  \\
    \midrule
    \multicolumn{7}{l}{\textbf{Non-competitive}}\\
    A   & 2x10 MHz in the 900 MHz band    & 3     & 0     & 1        &  1        & 1 \\
    \midrule
    \multicolumn{7}{l}{\textbf{Expenditure}}\\
       & In million DKK    &      & 820     &   485        &  1620        &107.6  \\
    \bottomrule
    \bottomrule
    \end{tabular}
    }

    \caption{Supply and auction outcome in the 2019 Danish spectrum auction}
    \begin{minipage}{.9\linewidth}
    \footnotesize{\emph{Notes:} R = reserve price in million DKK; the E lot came with a coverage obligation}
    \end{minipage}
    \label{tab:2019}
\end{table}

\section{Details of the 2021 Danish auction}\label{app:2021Danish}

The process was similar to the two previous auctions. Bidders first had the chance to obtain 2x10 MHz in the 2.1 GHz band with a coverage obligation for a reserve price of 0 (2.1-D lot). All bidders bought such a license. Table \ref{tab:2021} summarizes the supply and outcome.

There were two subsequent CMRAs. In the first CMRA, there were ten lots in the 1500 MHz band available: a single 25 MHz (lot 1.5-B) for a reserve price of DKK 10 million, eight 5 MHz lot (1.5-M) for a reserve price of DKK 10 million each, and another single 25 MHz block (1.5-T) for a reserve price of DKK 10 million. In the 2.1 GHz spectrum, there were six 2x5 MHz blocks (2.1-U) available for a reserve price of DDK 25 million each. In the 2.3 GHz band, there were two lots for 20 MHz available for a reserve price of DDK 25 million. In the 3.5 GHz band there were three categories of lots. First, there were three lots in the 3.5 GHz band available (3.5-D). The reserve price for such a lot was DDK 75 million. One such lot corresponds to 80 MHz in the 3.5 GHz spectrum and 400 MHz in the 26 GHz spectrum. Second, there was a single lot of 60 MHz (3.5-P) for a reserve price of DDK 25 million in the 3.5 GHz band with a leasing obligation. Third, there were nine 10 MHz lots (3.5-U) for a reserve price of DDK 25 million. The second CMRA was for the remaining lots in the 26 GHz band. In total there were 2850 MHz unpaired frequencies in the 26 GHz band. After subtracting the 1200 MHz sold in the first auction through the 3.5-D lots, there were 1650 MHz available (in lots of 200 MHz and 250 MHz) in the second CMRA. The reserve price was about DKK 5 million per lot. 

Each of the three bidders won a 2x10 MHz in the 2.1 GHz band with a coverage obligation for a reserve price of 0, two 2x5 MHz lots in the 2.1 GHz band, and a 3.5-D lot (80 MHz in the 3.5 GHz band and 400 MHz in the 26 GHz band). 

In addition, bidder A won 40 MHz in the 3.5 GHz band (four 3.5-U lots) in the first CMRA and 600 MHz in the 26 GHz band in the second CMRA. Bidder A’s total payment was DKK 540,525,000.

In addition to the above, bidder B won the 1.5-B lot, four 1.5-M lots, the two lots in the 2.3 GHz band, and 50 MHz in the 3.5 GHz band (five 3.5-U lots). In the second CMRA, bidder B won 850 MHz in the 26 GHz band. Bidder B’s total payment was DKK 794,685,000. 

In addition to the above, bidder C won the 1.5-T lot, four 1.5-M lots, and the 60 MHz in the 3.5 GHz spectrum with the leasing obligation. In the second CMRA, bidder C won 200 MHz in the 26 GHz band. Bidder C’s total payment was DKK 740,976,000.

We first look at the differences between bidders B and C. Bidder B won the two 2.3-U lots in the 2.3 GHz band (40 MHz in total), 10 MHz less in the 3.5 GHz band (but without the leasing obligation), and 650 MHz more in the 26 GHz band. Bidder B paid DKK 53,709,000 more than bidder C. Bidder B’s final assignment seems to dominate bidder C’s and cost only DKK 54 million more. Compare this number to the reserve price of DKK 100 million for the 2.3 GHz band alone. Hence, we suspect that bidder B used additional bids to win the large package (as under CMRA-truthful bidding with decreasing marginal values).

Next, we compare the outcomes of bidders A and B. Bidder B paid DKK 254 million more than bidder A and got the additional 45 MHz in the 1500 MHz band, 40 MHz in the 2300 MHz band, 10 MHz in the 3.5 GHz band, and 250 MHz in the 26 GHz band. The reserve price for the additional lots won by bidder B is DKK 180 million. Hence, bidder B paid DKK 74 million in excess of the reserve price.

Comparing bidders A and C, bidder C won 45 MHz in the 1500 MHz band while bidder A did not win any lot in this category. Bidder C won 20 MHz more in the 3.5 GHz band (but subject to the leasing obligation), and 400 MHz less in the 26 GHz band. Bidder A paid DKK 200 million less, however. The reserve price of the 45 MHz in the 1500 MHz band was DDK 50 million.

We conclude that it is likely that bidder B won with an additional bid. Due to the many prices, we cannot say whether bidders A and B won their headline demands or with additional bids. There is, however, no evidence for risk-free demand reduction.

\begin{table}
    \centering
    \footnotesize{
    \begin{tabular}{llrrrrr}
    \toprule
    \toprule
    Lot     & Description                   & Supply & R  & Bidder A  & Bidder B  & Bidder C\\
    \midrule
    \multicolumn{7}{l}{\textbf{CMRA}}\\
    1.5-B   & 25 MHz in the 1500 MHz band (bottom)   & 1     & 10                   &           & 1         &   \\
    1.5-M   & 5 MHz in the 1500 MHz band    & 8     & 10                            &           & 4         & 4 \\   
    1.5-T   & 25 MHz in the 1500 MHz band (top)  & 1     & 10                       &           &           & 1 \\
    2.1-U   & 2x5 MHz in the 2.1 GHz band   & 6     & 25                            & 2         & 2         & 2 \\
    2.3-U   & 20 MHz in the 2.3 GHz band    & 2     & 50                            &           & 2         &   \\
    3.5-D   & 80 MHz in 3.5 GHz + 400 MHz in 26 GHz & 3 & 75                        & 1         & 1         & 1 \\
    3.5-P   & 60 MHz in 3.5 GHz (leasing obligation) & 1 & 25                       &           &           & 1 \\
    3.5-U   & 10 MHz in the 3.5 GHz band    & 9     & 25                            & 4         & 5         &   \\
    26-U    & 200 MHz/250 MHz in the 26 GHz band & 8 & 5                            & 3         & 4         & 1 \\
    \midrule
    \multicolumn{7}{l}{\textbf{Non-competitive}}\\
    2.1-D    & 2x10 MHz in the 2.1 GHz band & 3 & 0                            & 1         & 1         & 1 \\
    \midrule
    \multicolumn{7}{l}{\textbf{Expenditure}}\\
        & In million DKK &  & 865                            & 541         & 795         & 741 \\
    \bottomrule
    \bottomrule
    \end{tabular}
    }
    \caption{Supply and auction outcome in the 2021 Danish spectrum auction}
    \begin{minipage}{.9\linewidth}
    \footnotesize{\emph{Note:} R = reserve price in million DKK}
    \end{minipage}
    
    \label{tab:2021}
\end{table}

\newpage
\section{Illustration of CMRA-truthful bidding}
\label{online appendix illustration CMRA truthful}
We now illustrate how the CMRA progresses under CMRA-truthful bidding in the setting of Section~\ref*{sec:strategic}; there are two bidders ($n=2$) and the single good ($m=1$) is perfectly divisible.

\subsection{Decreasing marginal values}\label{sec:truthful-decreasing-marginal-values}
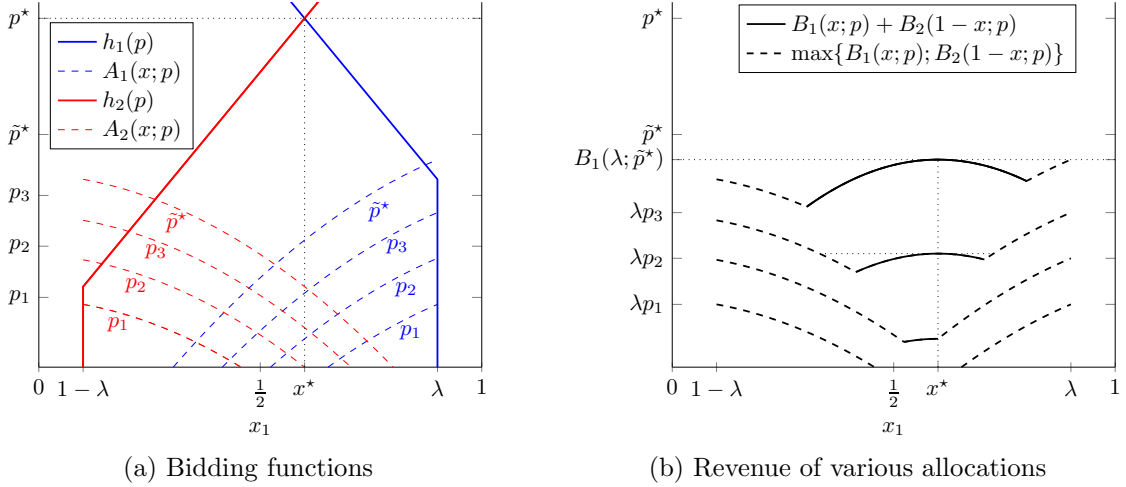
\begin{figure}
\centering
    \footnotesize{
	\subfloat[Bidding functions]{\label{fig:concave-bidding-functions}
\begin{tikzpicture}[scale=.85]

        \begin{axis}[
            domain=0:1,
            axis x line=left,
            x axis line style={-},
            xlabel = \(x_1\),
            samples=200,
            xtick={0, .1, .5, .6, .9, 1},
            xticklabels={$0$, $1-\lambda$, $\frac{1}{2}$, $x^\star$, $\lambda$, $1$},
            ytick={.13, 0.225379, .32, 0.433559, .65},
            yticklabels={$p_1$, $p_2$, $p_3$, $\tilde p^\star$, $p^\star$},
            ymax=.68,
            ymin=0,
            xmax=1,
            legend style={at={(0.025,.771)},anchor=west,legend columns=1},
            legend cell align={left}
        ]

        \addplot [blue, thick] (0,1.25)--(.9,0.35)--(.9,0);

        \addplot [blue, dashed, domain=0.652924:.9] {1.25 * x - .5 * x * x - .603};

        \addplot [red, thick] (.1,0)--(.1,.15)--(1,1.05);
        \addplot [red, dashed, domain=0.1:0.456458] {1.05 * (1-x) - .5 * (1-x) * (1-x) - 0.423};
        \legend{$h_1(p)$, $A_1(x;p)$,$h_2(p)$, $A_2(x;p)$}
        \addplot [blue, dashed, domain=0.523239:.9] {1.25 * x - .5 * x * x - 0.517159};
        \addplot [blue, dashed, domain=0.303468:.9] {1.25 * x - .5 * x * x - 0.333288};
        \addplot [blue, dashed, domain=0.414237:.9] {1.25 * x - .5 * x * x - 0.432}; %

        \addplot [red, thick] (.1,0)--(.1,.15)--(1,1.05);
        \addplot [red, dashed, domain=0.1:0.456458] {1.05 * (1-x) - .5 * (1-x) * (1-x) - 0.423};
        \addplot [red, dashed, domain=0.1:0.6] {1.05 * (1-x) - .5 * (1-x) * (1-x) - 0.34};
        \addplot [red, dashed, domain=0.1:0.8] {1.05 * (1-x) - .5 * (1-x) * (1-x) - 0.19};
        \addplot [red, dashed, domain=0.1:0.704718] {1.05 * (1-x) - .5 * (1-x) * (1-x) - 0.26645}; %

        \addplot [thin, dotted] (0,.65)--(.6,.65)--(.6,0);
        \addplot [thin, dotted] (.6,.65)--(1,.65);

        \node [red] at (.18,.08) {\footnotesize{$p_1$}};
        \node [red] at (.22,.152) {\footnotesize{$p_2$}};
        \node [red] at (.265,.215) {\footnotesize{$p_3$}};
        \node [red] at (.312,.27) {\footnotesize{$\tilde p^\star$}};
        \node [blue] at (.84,.065) {\footnotesize{$p_1$}};
        \node [blue] at (.83,.145) {\footnotesize{$p_2$}};
        \node [blue] at (.81,.225) {\footnotesize{$p_3$}};
        \node [blue] at (.77,.29) {\footnotesize{$\tilde p^\star$}};

        \end{axis}
\end{tikzpicture}
}\qquad
\subfloat[Revenue of various allocations]{\label{fig:concave-revenue}
\begin{tikzpicture}[scale=.85]

        \begin{axis}[
            domain=0.1:.9,
            axis x line=left,
            x axis line style={-},
            xlabel = \(x_1\),
            samples=200,
            xtick={0, .1, .5, .6, .9,1},
            xticklabels={0,$1-\lambda$, $\frac{1}{2}$, $x^\star$, $\lambda$,1},
            ytick={0.117, 0.2028411, .288, 0.386712, 0.433559,.65},
            yticklabels={$\lambda p_1$, $\lambda p_2$, $\lambda p_3$, $B_1(\lambda;\tilde p^\star)$,$\tilde p^\star$, $p^\star$},
            ymax=.68,
            ymin=0,
            xmin=0,
            xmax=1,
            legend style={at={(0.15,.9)},anchor=west,legend columns=1},
            legend cell align={left}
        ]
        \addplot [thick, domain=0.303468:0.8] {max(0,1.05 * (1-x) - .5 * (1-x) * (1-x) - 0.19) + max(0,1.25 * x - .5 * x * x - 0.333288)};

        \addplot [thick, dashed,domain=0.652924:.9] {1.25 * x - .5 * x * x - .603};
        \addplot [thick, dashed, domain=0.1:0.456458] {1.05 * (1-x) - .5 * (1-x) * (1-x) - 0.423};
        \legend{$B_1(x;p) + B_2(1-x;p)$, $\max\{B_1(x;p) ; B_2(1-x;p)\}$}

        \addplot [thick,dashed] {max(0,1.05 * (1-x) - .5 * (1-x) * (1-x) - 0.34)+max(0,1.25 * x - .5 * x * x - 0.517159)};

        \addplot [thick, dashed,, domain=0.1:.9] {max(0,1.05 * (1-x) - .5 * (1-x) * (1-x) - 0.19) + max(0,1.25 * x - .5 * x * x - 0.333288)};

        \addplot [thick, dashed, domain=0.1:.9] {max(0,1.05 * (1-x) - .5 * (1-x) * (1-x) - 0.26645) + max(0,1.25 * x - .5 * x * x - 0.432)};
        
        \addplot [thick, domain=0.303468:0.8] {max(0,1.05 * (1-x) - .5 * (1-x) * (1-x) - 0.19) + max(0,1.25 * x - .5 * x * x - 0.333288)};
        \addplot [thick, domain=0.414237:0.704718] {max(0,1.05 * (1-x) - .5 * (1-x) * (1-x) - 0.26645) + max(0,1.25 * x - .5 * x * x - 0.432)};
        \addplot [thick, domain=0.523239:0.6] {max(0,1.05 * (1-x) - .5 * (1-x) * (1-x) - 0.34)+max(0,1.25 * x - .5 * x * x - 0.517159)};

        \addplot [thin, dotted] (0.332753,.21155)--(0.725214,0.21155);
        \addplot [thin, dotted] (0,0.386712)--(1,0.386712);
        \addplot [thin, dotted] (0.6,0)--(.6,0.386712);
        \end{axis}
\end{tikzpicture}     

}
	}
    \caption{CMRA-truthful bidding with decreasing marginal values.}
    \label{fig:truthful-decreasing-marginal-values}
    \begin{minipage}{.9\linewidth}
    \footnotesize{\emph{Note:} We depict $\max\{B_1(x;p),B_2(1-x;p)\}$ only if the allocation $(x,1-x)$ has not received bids from both bidders.}
    \end{minipage}
\end{figure}

Figure~\ref{fig:truthful-decreasing-marginal-values} illustrates the results of Section~\ref*{sec:truthful} in the symmetric-caps case. Figure~\ref{fig:concave-bidding-functions} shows the headline demands and additional bids at various clock prices. Bidder 1 is stronger, so the efficient share is $x_1^\star>\frac{1}{2}$ while bidder 2's efficient share is $x_2^\star<\frac{1}{2}$. Solid lines are headline demands $h_i(p)$ and dashed lines are truthful additional bids $A_i(x;p)$ as in Eq.~(\ref*{eq:additional}) in the main text). Figure~\ref{fig:concave-revenue} depicts the respective revenue from feasible allocations. The solid line $B_1(x;p) + B_2(1-x;p)$ shows revenue for allocations in which a bid of each bidder is accepted since this is required by the CMRA closing rule (recall that bids are $-\infty$ for shares that bidders do not bid on). The dashed line is $\max\{B_1(x;p),B_2(1-x;p)\}$ for allocations that do not receive non-negative bids from both bidders: this is revenue that can be obtained by accepting only one bidder's bid. 

Let us consider how the bids and allocations change as the clock price increases. As a benchmark, consider a simple clock auction (or a CMRA with clock-truthful bidding). We simply increase the clock price and follow the headline demands in solid lines in Figure~\ref{fig:concave-bidding-functions}. The auction ends at clock price $p^\star$ with market clearing.

Under CMRA-truthful bidding, both bidders submit headline demands and additional bids. When the clock price $p$ is low, only quantities close to $\lambda$ receive additional bids. 
\begin{itemize}
    \item \emph{Clock price $p_1$}.
At this clock price, each bidder's headline demand is $\lambda$, yielding surplus $U_i(\lambda) - \lambda p_1$. Recall that the additional bids are given by Eq.~(\ref*{eq:additional}) in the main text. At $p_1$, the additional bids range from $\lambda p_1$ (for a quantity $\lambda$) to zero (for a smaller quantity that keeps the bidder indifferent). At $p_1$, there is no feasible allocation that receives bids from both bidders (the dashed lines do not intersect in Figure~\ref{fig:concave-revenue}). The auction continues as it is not possible to accept a bid by each bidder in the revenue-maximizing allocation.

\item \emph{Clock price $p_2$}. This is the lowest price at which both bidders bid on their respective efficient quantities. Bidder 1 submits a strictly positive additional bid on $x_1^\star$ at clock price $p_2$, while bidder 2 submits an additional bid of 0 on $x_2^\star$. From now on the efficient allocation can in principle be allocated as it receives bids from both bidders. Figure~\ref{fig:concave-revenue} reveals, however, that the efficient allocation is not revenue-maximizing. Bidder 1's headline demand is still $\lambda$, and allocating $\lambda$ to bidder 1 raises a revenue of $\lambda p_2$. Observe that bidder 2 bids less than $\lambda p_2$ on $\lambda$ because marginal values are decreasing and because $\lambda$ is not the headline demand. As bidder 2 does not bid on $1-\lambda$ at clock price $p_2$, the revenue-maximizing allocation features only bids by one bidder and the auction continues.

\item \emph{Clock price $p_3$}. Both bidders have now raised their additional bids on their respective efficient share. As we can see in Figure~\ref{fig:concave-revenue}, the efficient allocation $x^\star$ locally maximizes revenue. However, $x^\star$ does not yield a global revenue maximum as bidder 1's bid on $\lambda$ leads to a higher revenue of $\lambda p_3$.

\item{\emph{Clock price} $\tilde p^\star$.} The additional bids are now sufficiently high so that the efficient allocation $x^\star$ is revenue-maximizing. The auction ends at $\tilde p^\star$ at which 
\begin{equation}
    B_1(x^\star_1;\tilde p^\star) + B_2(x^\star_2;\tilde p^\star) = \max_i B_i(\lambda;\tilde p^\star).
\end{equation}
Note that at price $\tilde p^\star$ bidder 2 does not yet bid on $1-\lambda$, so $(\lambda,1-\lambda)$ is not a feasible allocation. 
Revenue is lower than $\tilde p^\star$ and lower than $p^\star$ (Fig.~\ref{fig:concave-revenue}).
\end{itemize}

\subsection{Non-decreasing marginal values}
We now consider non-decreasing marginal values. Since no competitive equilibrium needs to exist, we prove a revenue comparison between clock-truthful and CMRA-truthful bidding.

\begin{proposition}
    Let the marginal values be non-decreasing.
	\begin{enumerate}[label=(\roman*)]
	    \item Clock-truthful bidding in the CMRA leads to excess supply. The clock ends at clock price $p = \min_i U_i(\lambda_i)/\lambda_i$. The final auction allocation is inefficient.
	    \item If $\lambda_1=\lambda_2$, then ex-post revenue under CMRA-truthful bidding is lower than under clock-truthful bidding; if $\lambda_1>\lambda_2$, then the ex-post revenue comparison is ambiguous.
	\end{enumerate}
	\label{theorem:truthful-non-decreasing-marginal-values}
\end{proposition}

\begin{proof}
    The clock-truthful demand is $\lambda_i$ until the price reaches $U_i(\lambda_i)/\lambda_i$. For higher prices demand equals 0. Hence, the clock ends at price $\min U_i(\lambda_i)/\lambda_i$ and it does so with excess supply. Due to positive marginal values for all shares below $\lambda_i$, the efficient allocation does not feature excess supply. The outcome is inefficient.
    
    Let $\lambda_1=\lambda_2=\lambda$. Under clock-truthful bidding, the clock ends at clock price $\min_i U_i(\lambda)/\lambda$. Under CMRA-truthful bidding the auction ends at a lower clock price, namely $\min_i p_i^f$. As ex-post revenue is $\lambda$ times the final clock price in both cases, revenue is lower under CMRA-truthful bidding due to the lower final clock price.

    The following numerical example proves that revenue can also be higher under CMRA-truthful bidding. Let $\li=\frac{7}{8}$, $\lj=\frac{6}{8}$, $U_1(\li)=21$, $U_1(1-\lj) =1$, $U_2(\lj)=\frac{39}{2}$, and $U_2(1-\li)=\frac{1}{2}$.\footnote{It is straightforward to check that the utility functions are consistent with non-decreasing marginal values. The difference $\li-(1-\lj)=\lj-(1-\li)=\frac{5}{8}$ and $U_1(\li)-U_1(1-\lj)=20>19=U_2(\lj)-U_2(1-\li)$.} The allocation $(\li,1-\li)$ is efficient. Under clock-truthful bidding, the CMRA ends at clock price $\min_i U_i(\lambda_i)/\lambda_i = U_1(\li)/\li=24 < 26 = U_2(\lj)/\lj$. Consider CMRA-truthful bidding. Since $(\li,1-\li)$ is efficient, Theorem 1 implies that the CMRA must end at clock price $p_2^f=76/3$. Revenue is $\pjf\li=\frac{133}6$, which is more than the revenue under clock-truthful bidding: $\lj U_1(\li)/\li=18.$
\end{proof}

Figure~\ref{fig:truthful-increasing-marginal-values} illustrates Proposition~\ref{theorem:truthful-non-decreasing-marginal-values} for symmetric caps. As before, Figure~\ref{fig:convex-bidding-functions} shows the headline demands and the additional bids, while Figure~\ref{fig:convex-revenue} shows revenue under different allocations.

Once again, let us first consider the outcome of a clock auction or of clock-truthful bidding in the CMRA. Figure~\ref{fig:convex-bidding-functions} shows that, due to increasing marginal values, bidder $i$'s clock-truthful headline demand is $\lambda$ for $p\le U_i(\lambda)/\lambda$ and 0 for higher prices. Hence, the auction ends at clock price $p=\min_i U_i(\lambda)/\lambda = U_2(\lambda)/\lambda$. As bidder 2 drops demand to 0 at price $U_2(\lambda)/\lambda$, the auction ends with excess supply of $1-\lambda$. Bidder 1 wins quantity $\lambda$ and the revenue is $U_2(\lambda)$.
Note that in this case, clock-truthful bidding in the CMRA is equivalent to a VCG auction restricted to selling $\lambda$ as a bundle (i.e., a second-price auction for $\lambda$), so the revenue is lower than in the (unrestricted) VCG auction.\footnote{We are grateful to an anonymous referee for this observation.}

\begin{figure}
\centering
    \footnotesize{
	\subfloat[Bidding functions]{\label{fig:convex-bidding-functions}
\begin{tikzpicture}[scale=.85]

        \begin{axis}[
            domain=0:1,
            axis x line=left,
            x axis line style={-},
            xlabel = \(x_1\),
            samples=200,
            xtick={0, .1, .5, .9, 1},
            xticklabels={$0$, $1-\lambda$, $\frac{1}{2}$, $\lambda$, $1$},
            ytick={.13, .26859890446000095, .35, 0.426667, 0.471111, 0.571111},
            yticklabels={$p_1$, $p_2$, $p_3$, $p_2^f$, $\frac{U_2(\lambda)}\lambda$, $\frac{U_1(\lambda)}\lambda$},
            ymax=.6,
            ymin=0,
            xmax=1,
            xmin=0,
            legend style={at={(0.15,.8)},anchor=west,legend columns=1},
            legend cell align={left}
        ]

        \addplot [blue, thick] (.9,0.571111)--(.9,0);

        \addplot [blue, dashed, domain=0.75918:.9] { x/2 + (x - .5) * (x - .5) * (x - .5) - 0.397};

        \addplot [red, thick] (.1,0)--(.1,0.471111); %
        \addplot [red, dashed, domain=0.1:0.264961] {.4 * (1-x) + (.5 - x) * (.5 - x) * (.5 - x) - 0.307}; %
        \legend{$h_1(p)$, $A_1(x;p)$,$h_2(p)$, $A_2(x;p)$}

        \addplot [blue, dashed, domain=0.544348:.9] {x/2 + (x - .5) * (x - .5) * (x - .5) - 0.272261}; %
        \addplot [blue, dashed, domain=0.5:.9] {x/2 + (x - .5) * (x - .5) * (x - .5) - 0.199}; %
        \addplot [blue, dashed, domain=0.398:.5] {x/2  - 0.199}; %
        \addplot [blue, dashed, domain=0.5:.9] {x/2 + (x - .5) * (x - .5) * (x - .5) - 0.13}; %
        \addplot [blue, dashed, domain=0.26:.5] {x/2 - 0.13}; %

        \addplot [red, dashed, domain=0.5:0.541276] {.4 * (1-x) + (.5 - x) * (.5 - x) * (.5 - x) - 0.182261}; %
        \addplot [red, dashed, domain=0.1:0.5] {.4 * (1 - x) + ( .5 - x ) * ( .5 - x) * ( .5 - x) - 0.182261}; %

        \addplot [red, dashed, domain=0.5:0.7275] {.4 * (1-x) - 0.109}; %
        \addplot [red, dashed, domain=0.1:0.5] {.4 * (1 - x) + ( .5 - x ) * ( .5 - x) * ( .5 - x) - 0.109}; %

        \addplot [red, dashed, domain=0.5:0.9] {.4 * (1-x) - 0.04}; %
        \addplot [red, dashed, domain=0.1:0.5] {.4 * (1 - x) + ( .5 - x ) * ( .5 - x) * ( .5 - x) - 0.04}; %

        \node [red] at (.17,.04) {\footnotesize{$p_1$}};
        \node [red] at (.222,.13) {\footnotesize{$p_2$}};
        \node [red] at (.265,.175) {\footnotesize{$p_3$}};
        \node [red] at (.32,.275) {\footnotesize{$p_2^f$}};
        \node [blue] at (.88,.065) {\footnotesize{$p_1$}};
        \node [blue] at (.85,.16) {\footnotesize{$p_2$}};
        \node [blue] at (.82,.21) {\footnotesize{$p_3$}};
        \node [blue] at (.725,.285) {\footnotesize{$ p^f_2$}};

        \end{axis}
\end{tikzpicture}
}\qquad
\subfloat[Revenue of various allocations]{\label{fig:convex-revenue}
\begin{tikzpicture}[scale=.85]

        \begin{axis}[
            domain=0.1:.9,
            axis x line=left,
            x axis line style={-},
            xlabel = \(x_1\),
            samples=200,
            xtick={0, .1, .5, .9, 1},
            xticklabels={0, $1-\lambda$, $\frac{1}{2}$, $\lambda$,1},
            ytick={0.117, 0.241739014, 0.315, 0.3840003},
            yticklabels={$\lambda p_1$, $\lambda p_2$, $\lambda p_3$, $B_1(\lambda;\tilde p^\star)$},
            ymax=.6,
            ymin=0,
            xmin=0,
            xmax=1,
            legend style={at={(0.15,.9)},anchor=west,legend columns=1},
            legend cell align={left}
        ]
        \addplot [thick, domain=.5:.9] {.4 * (1-x) - 0.04 + x/2 + (x - .5) * (x - .5) * (x - .5) - 0.13}; 
        \addplot [dashed,thick,domain=0.652924:.9] {x/2 + (x - .5) * (x - .5) * (x - .5) - 0.397};
        \legend{$B_1(x;p) + B_2(1-x;p)$, $\max\{B_1(x;p);B_2(1-x;p)\}$}

        \addplot [dashed, thick, domain=0.1:0.264961] {.4 * (1-x) + (.5 - x) * (.5 - x) * (.5 - x) - 0.307}; 
        \addplot [thick, dashed, domain=0.544348:.9] {x/2 + (x - .5) * (x - .5) * (x - .5) - 0.272261}; %
        \addplot [thick, dashed, domain=0.5:0.541276] {.4 * (1-x) + (.5 - x) * (.5 - x) * (.5 - x) - 0.182261}; %
        \addplot [thick, dashed, domain=0.1:0.5] {.4 * (1 - x) + ( .5 - x ) * ( .5 - x) * ( .5 - x) - 0.182261}; %

        \addplot [thick,dashed, domain=0.1:0.398] {.4 * (1 - x) + ( .5 - x ) * ( .5 - x) * ( .5 - x) - 0.109}; 
        \addplot [thick, domain=0.398:.5] {.4 * (1 - x) + ( .5 - x ) * ( .5 - x) * ( .5 - x) - 0.109 + x/2 - 0.199}; 
        \addplot [thick, domain=.5:.7275] {.4 * (1-x) - 0.109 + x/2 + (x - .5) * (x - .5) * (x - .5) - 0.199}; 
        \addplot [thick, dashed, domain=.7275:.9] { x/2 + (x - .5) * (x - .5) * (x - .5) - 0.199}; 

        \addplot [thick, dashed, domain=0.1:0.26] {.4 * (1 - x) + ( .5 - x ) * ( .5 - x) * ( .5 - x) - 0.04}; 
        \addplot [thick, domain=0.26:.5] {.4 * (1 - x) + ( .5 - x ) * ( .5 - x) * ( .5 - x) - 0.04 + x/2 - 0.13}; 

        \end{axis}
\end{tikzpicture}     

}
	}
    \caption{CMRA-truthful bidding with non-decreasing marginal values}
    \label{fig:truthful-increasing-marginal-values}
    \begin{minipage}{.9\linewidth}
    \footnotesize{\emph{Note:} We depict $\max\{B_1(x;p),B_2(1-x;p)\}$ only if the allocation $(x,1-x)$ has not received bids from both bidders.}
    \end{minipage}
\end{figure}
Under CMRA-truthful bidding, bidders' headline demands are as under clock-truthful bidding but they also submit additional bids. For low clock prices, bidders submit few additional bids, but as the clock price rises, bidders increase their additional bids both on the intensive and extensive margins. 
\begin{itemize}
    \item \emph{Clock price $p_1$}. There is no feasible allocation that receives non-negative bids from both bidders. Hence, the auction continues.
    \item \emph{Clock price $p_2$}. There are now feasible allocations that receive non-negative bids from both bidders. These allocations are not revenue-maximizing as $B_i(\lambda;p)=\lambda p$ yields higher revenue.
    \item \emph{Clock price $p_3$}. Feasible allocations that receive non-negative bids from both bidders are still not revenue-maximizing. Bidder 1's marginal values (and additional bids) are higher and non-decreasing, so allocating more to bidder~1 increases revenue.
    \item \emph{Clock price $p^f_2$}. At this price, the weaker bidder 2 places an additional bid of 0 on $1-\lambda$. More generally, there is a \emph{final price} $p_i^f$ at which bidder $i$ bids 0 on $1-\lambda$ as this bidder is indifferent between winning $\lambda$ for a payment of $ p_i^f\lambda$ and winning $1-\lambda$ for free. The indifference condition $U_i(\lambda) - p^f_i\lambda = U_i(1-\lambda)$ transforms to
\begin{equation*}
    p^f_i = \frac{U_i(\lambda_i) - U_i(1-\lambda_j)}{\lambda_i}.
\end{equation*}
With bidder~2's additional bid, it is now possible to accept a bid by each bidder in the revenue-maximizing allocation $(\lambda,1-\lambda)$ (Figure~\ref{fig:convex-revenue}). Therefore, the CMRA ends in market-clearing. The revenue is $\lambda p_2^f$, which is lower than $\lambda  U_2(\lambda)/\lambda$.
\end{itemize}

\end{document}